%% file: main.tex
\documentclass[fleqn,10pt]{wlscirep}
\usepackage[utf8]{inputenc}
\usepackage[T1]{fontenc}
\usepackage{amsthm,amsmath,amssymb}
\usepackage{mathrsfs}
\usepackage{graphicx}
\usepackage{multirow}
\DeclareMathAlphabet{\mathcal}{OMS}{cmsy}{m}{n}
\usepackage{colortbl}
\definecolor{bk}{HTML}{A4E2C6}

\title{
Deep Learning-based Diffusion Tensor Cardiac Magnetic Resonance Reconstruction: A Comparison Study
}

\author[1,2]{Jiahao Huang}
\author[1,2]{Pedro F. Ferreira}
\author[1,3]{Lichao Wang}
\author[1,2]{Yinzhe Wu}
\author[4]{Angelica I. Aviles-Rivero}
\author[4]{Carola-Bibiane Sch{\"o}nlieb}
\author[1,2]{Andrew D. Scott}
\author[1,2]{Zohya Khalique}
\author[1,2]{Maria Dwornik}
\author[1,2]{Ramyah Rajakulasingam}
\author[1,2]{Ranil De Silva}
\author[1,2]{Dudley J. Pennell}
\author[1,2,*]{Sonia Nielles-Vallespin}
\author[1,2,*]{Guang Yang}
\affil[1]{National Heart and Lung Institute, Imperial College London, London, United Kingdom}
\affil[2]{Cardiovascular Research Centre, Royal Brompton Hospital, London, United Kingdom}
\affil[3]{Department of Computing, Imperial College London, London, United Kingdom}
\affil[4]{Department of Applied Mathematics and Theoretical Physics, University of Cambridge, Cambridge, United Kingdom}

\affil[*]{Co-last senior authors. Send correspondence to \{j.huang21,g.yang\}@imperial.ac.uk}


\begin{abstract}
In vivo cardiac diffusion tensor imaging (cDTI) is a promising Magnetic Resonance Imaging (MRI) technique for evaluating the micro-structure of myocardial tissue in the living heart, providing insights into cardiac function and enabling the development of innovative therapeutic strategies. 
However, the integration of cDTI into routine clinical practice is challenging due to the technical obstacles involved in the acquisition, such as low signal-to-noise ratio and long scanning times. 
In this paper, we investigate and implement three different types of deep learning-based MRI reconstruction models for cDTI reconstruction. We evaluate the performance of these models based on reconstruction quality assessment and diffusion tensor parameter assessment. 
Our results indicate that the models we discussed in this study can be applied for clinical use at an acceleration factor (AF) of $\times 2$ and $\times 4$, with the D5C5 model showing superior fidelity for reconstruction and the SwinMR model providing higher perceptual scores. There is no statistical difference with the reference for all diffusion tensor parameters at AF $\times 2$ or most DT parameters at AF $\times 4$, and the quality of most diffusion tensor parameter maps are visually acceptable. 
SwinMR is recommended as the optimal approach for reconstruction at AF $\times 2$ and AF $\times 4$. 
However, we believed the models discussed in this studies are not prepared for clinical use at a higher AF.
At AF $\times 8$, the performance of all models discussed remains limited, with only half of the diffusion tensor parameters being recovered to a level with no statistical difference from the reference. Some diffusion tensor parameter maps even provide wrong and misleading information. 
\end{abstract}

\begin{document}

\flushbottom
\maketitle




\section*{Introduction}

In vivo cardiac diffusion tensor (DT) imaging (cDTI) is an emerging Magnetic Resonance Imaging (MRI) technique that has the potential to describe the micro-structure of myocardial tissue in the living heart. 
The diffusion of water molecules occurs anisotropically due to the restrictions imposed by the micro-structure of the myocardium, which can be approximated by fitting three-dimensional (3D) tensors with a specific shape and orientation in cDTI. 
Various parameters can be derived from the DT, including mean diffusivity (MD) and fractional anisotropy (FA), which are crucial indices that can indicate the structural integrity of myocardial tissues. The helix angle (HA) signifies local cell orientations, while the second eigenvector (E2A) represents the average sheetlet orientation~\cite{Ferreira2014_invivo}. 
The development of cDTI provides insights into the myocardial micro-structure and offers new perspectives on the elusive connection between cellular contraction and macroscopic cardiac function~\cite{Ferreira2014_invivo,Sonia2017_Assessment}. Furthermore, it presents opportunities for novel assessments of the myocardial micro-structure and cardiac function, as well as the development and evaluation of innovative therapeutic strategies~\cite{Khalique2020_Diffusion}. 

Despite the numerous advantages, there are still significant technical obstacles that must be overcome to integrate cDTI into routine clinical practice.
For the calculation of the DT, diffusion-weighted images (DWIs) with diffusion encoding in at least six distinct directions need to be collected. Due to the movement derived from the heart beat and human breath, in vivo cDTI exploits single-shot encoding acquisition for repetitive fast scanning, e.g., single-shot echo planar imaging (SS-EPI) or spiral diffusion-weighted imaging~\cite{Basser1995_Inferring}. The utilisation of these single-shot encoding acquisitions, which lead to low signal-to-noise (SNR) images, usually requires multiple repetitions to enhance the accuracy of the DT estimation\cite{Scott2016_effects,Ma2018_Accelerated}. 
Each repetition necessitates an additional breath-hold for the patient when using breath-hold acquisitions, which significantly increases the total scanning time and leads to uncomfortable patient experience.

Numerous studies have been proposed to accelerate cDTI technique, which can be mainly categorised as 
1) reducing the total amount of DWIs used for the calculation of the DT;
2) general fast DWIs by \textit{k}-space undersampling and reconstruction using compressed sensing (CS) or deep learning techniques. 
This study focuses on the second strategy.

Deep learning has emerged as a powerful technique for image analysis, capitalising on the non-linear and complex nature of networks through supervised or unsupervised learning, and has found widespread applications in medical image researches~\cite{Shen2017_Deep}.
Deep learning-based MRI reconstruction~\cite{Chandra2012_deep,Chen2022_AI} has gained significant attention, leveraging its capable of learning complex and hierarchical representations from large MRI datasets~\cite{Zbontar2018_fastMRI}. 

In this work, we investigate the application of deep learning-based methods for cDTI reconstruction. 
We explore and implement three different types of deep learning-based models on cDTI datasets with acceleration factor (AF) of $\times 2$, $\times 4$ and $\times 8$.
These models include
a Convolutional Neural Network (CNN)-based unrolling method, i.e., D5C5~\cite{Schlemper2017_D5C5}, 
a CNN-based and conditional Generative Adversarial Network (GAN)-based non-unrolling method, i.e., DAGAN~\cite{Yang2018_DAGAN}, 
and a Transformer-based non-unrolling methods, i.e., SwinMR~\cite{Huang2022_SwinMR}.
The performance of these three models are evaluated by the reconstruction quality assessment and the DT parameters assessment. 

Our experiments demonstrate that the models discussed in this paper can be applied for clinical use at AF $\times 2$ and AF $\times 4$, since the both the reconstruction of DWIs and DT parameters reach satisfactory levels. 
Among these models, D5C5 shows superior fidelity for the reconstruction, while SwinMR provides results with higher perceptual scores. 
There is no statistical difference with the reference for all the DT parameters at AF $\times 2$ or most of the DT parameters at AF $\times 4$. 
The quality of most the DT parameter maps we considered are visually acceptable.
Considering various factors, SwinMR is recommended as the optimal approach for the reconstruction with AF $\times 2$ and AF $\times 4$.

However at AF $\times 8$, the performance of these three models, including the best-performing SwinMR, is still limited.
The reconstruction quality is not unsatisfactory due to the artefact remaining and the noisy (DAGAN) or `fake' (SwinMR) estimation. Only half of the DT parameters can be recovered to a level that is no statistical difference with the reference. Some DT parameter maps even provide wrong and misleading information, which is unacceptable and dangerous for clinical use.

\section*{Related Works}

\subsection*{Diffusion Tensor MRI Acceleration}

A major drawback of DTI is its extended scanning time, as it requires multiple DWIs with varying b-value and diffusion gradient directions to calculate the DT. 
In theory, the estimation of the DT requires only six DWIs with different diffusion gradient directions and one reference image.
Practically for cDTI, a considerable number of cardiac DWIs and multiple averages are typically required to enhance the accuracy of DT estimation, due to the inherently low SNR of single-shot acquisitions.

Strategies to accelerate the DTI technique have been explored. One technical route aims to reduce the number of DWIs required for the DT estimation~\cite{Ferreira2022_Accelerating,Karimi2022_Diffusion,Aliotta2021_Extracting,Li2021_SuperDTI,Tian2020_DeepDTI,Phipps2021_Accelerated,Tanzer2022_Faster}, which can be further categorised into three sub-class.

1) Learn a direct mapping from reduced repetition (or gradient direction) of DWIs, to the DT or DT parameter maps.
Ferreira~\textit{et~al.}~\cite{Ferreira2022_Accelerating} proposed a U-Net-based method for cDTI acceleration, which directly estimates the DT, using DWIs collected within one breath-hold, instead of solving a conventional linear-least-square (LLS) tensor fitting. 
Karimi~\textit{et~al.}~\cite{Karimi2022_Diffusion} introduced a Transformer-based model with coarse-and-fine strategy to provide accuracy estimation of the DT, using only six diffusion-weighted measurements. 
Aliotta~\textit{et~al.}~\cite{Aliotta2019_Highly} proposed a neural network for brain DTI, namely DiffNet, which estimated MD and FA maps directly from diffusion-weighted acquisitions with as few as three diffusion-encoding directions. 
They further improved their method by combining a parallel U-Net for slice-to-slice mapping and a multi-layer perceptron for pixel-to-pixel mapping~\cite{Aliotta2021_Extracting}. 
Li~\textit{et~al.}~\cite{Li2021_SuperDTI} developed a CNN-based model for brain DTI, i.e., SuperDTI, to generate FA, MD and directionally encoded color maps with as few as six diffusion-weighted acquisitions.

2) Enhance DWIs (denoising). This kind of methods usually apply only a small amount of enhanced images to achieve comparable estimation results with the results via standard protocol.
Tian~\textit{et~al.}~\cite{Tian2020_DeepDTI} developed a novel DTI processing framework, entitled DeepDTI, that minimised the required data for DTI to six diffusion-weighted images. The core idea of this framework was to use a CNN, which takes a non-diffusion-weighted (b0) image, six DWIs as well as a anatomical (T1- or T2-weighted) image as input, to produce high-quality b0 images and six DWIs. 
Phipps~\textit{et~al.}~\cite{Phipps2021_Accelerated} applied a denoising CNN to enhance the quality of b0 images and corresponding DWIs for cDTI. 

3) Refine the DT quality. 
T{\"a}nzer~\textit{et~al.}~\cite{Tanzer2022_Faster} proposed a GAN-based Transformer, to directly enhance the quality of the DT that was calculated with reduce amount of DWIs in a end-to-end manner.

Another technical route follows the general DWIs acceleration by \textit{k}-space undersampling and reconstruction~\cite{Zhu2017_Direct,Chen2018_Angular,Huang2019_Accelerating,Teh2020_Improved}.
Zhu~\textit{et~al.}~\cite{Zhu2017_Direct} directly estimated the DT from highly undersampled \textit{k}-space data.
Chen~\textit{et~al.}~\cite{Chen2018_Angular} incorporated the joint sparsity prior of different DWIs with the L1-L2 norm and the DT's smoothness using the total variational (TV) semi-norm to efficiently expedite DWI reconstruction.
Huang~\textit{et~al.}~\cite{Huang2019_Accelerating} utilised a local low-rank model and 3D TV constraints to reconstruct the DWIs from undersampled \textit{k}-space measurements.
Teh~\textit{et~al.}~\cite{Teh2020_Improved} introduced a directed TV-based method for DWI images reconstruction, applying the information on the position and orientation of edges in the reference image.

In addition to these major technical routes, Liu~\textit{et~al.}~\cite{Liu2023_Accelerated} explored the deep learning-based image synthetics for the inter-directional DWIs generation. The true b0 and 6 DWIs were concatenated with the generated data and passed to the CNN-based tensor fitting network.

\subsection*{Deep Learning-Based Reconstruction}

The aim of MRI reconstruction is to recover the ground truth image $x$ from the undersampled \textit{k}-space measurement $y$, which is mathematically described as an inverse problem:
\begin{equation}\label{eq:mri_reverse}
\begin{aligned}
x = \operatorname{arg}\min_{x} \frac{1}{2}|| \mathcal{A} x - y ||_2^2 + \lambda \mathcal{R}(x),
\end{aligned}
\end{equation}

\noindent in which the degradation matrix $\mathcal{A}$ can be further presented as the combination of the undersampling trajectory $\mathcal{M}$, Fourier transform $\mathcal{F}$ and coil sensitivity maps $\mathcal{S}$. $\lambda$ is the coefficient that balances regularisation term $\mathcal{R}(x)$.

Deep learning technique has been widely used for MRI reconstruction. Based on the association with traditional iterative CS algorithms, deep learning-based MRI reconstruction methods can be categorised into 1) unrolling-based models~\cite{Yang2016_DeepADMMNet,Schlemper2017_D5C5,Aggarwal2019_MoDL} and 2) non-unrolling-based models~\cite{Yang2018_DAGAN,Huang2022_SwinMR}.

Unrolling-based models usually integrate neural networks with traditional CS algorithms, simulating the iterative reconstruction algorithms through learnable iterative blocks~\cite{Chen2022_AI}.
Yang~\textit{et~al.}~\cite{Yang2016_DeepADMMNet} reformulated an Alternating Direction Method of Multipliers (ADMM) algorithm to a multi-stage deep architecture, namely Deep-ADMM-Net, for MRI reconstruction, of which each stage corresponds to an iteration in traditional ADMM algorithm.
Some unrolling-based models improved Eq~\eqref{eq:mri_reverse} with a deep learning-based regulariser~\cite{Schlemper2017_D5C5,Aggarwal2019_MoDL}, which can be formulated as:
\begin{equation}\label{eq:mri_reverse_dl}
\begin{aligned}
x = \operatorname{arg}\min_{x} \frac{1}{2}|| \mathcal{A} x - y ||_2^2 + \lambda || x - f_{\theta} (x_u) ||_2^2, \quad \text{s.t. } x_u = \mathcal{F}^{-1} y,
\end{aligned}
\end{equation}

\noindent in which $f_{\theta}(\cdot)$ is a deep neural network and $x_u$ is the undersampled zero-filled images (ZF).
Schlemper~\textit{et~al.}~\cite{Schlemper2017_D5C5} designed a deep cascade of CNNs for cardiac cine reconstruction, in which a spatio-temporal correlations can be also efficiently learned via the data sharing approach.
Aggarwal~\textit{et~al.}~\cite{Aggarwal2019_MoDL} proposed a model-based deep learning methods, namely MoDL, which exploited a CNN-based regularisation prior for MRI reconstruction.

Non-unrolling-based models usually train an deep learning-based function $f_{\theta}(\cdot)$ that maps the undersampled \textit{k}-space measurement $y$ or zero-filled images $x_u$ to estimate fully-sampled images $\hat x_u$ or its residual in an end-to-end manner, which can be formulated as $\hat x_u = f_{\theta}(x_u)$ or $\hat x_u = f_{\theta}(x_u) + x_u$.
Yang~\textit{et~al.}~\cite{Yang2018_DAGAN} proposed a de-alising Generative Adversarial Networks for MRI reconstruction, in which the U-Net-based generator produced the estimated fully-sampled MRI images in an end-to-end manner.
Feng~\textit{et~al.}~\cite{Feng2021_Task} exploited the task-specific novel cross-attention and designed a end-to-end Transformer-based model for jointly MRI reconstruction and super-resolution.
Huang~\textit{et~al.}~\cite{Huang2022_SwinMR} proposed a Swin Transformer-based model, namely SwinMR, for end-to-end MRI reconstruction, and they further explored the combination of Swin Transformer and GAN for the edge and texture preservation in MRI reconstruction~\cite{Huang2022_STGAN}.

Deep learning community constantly provides a wide range of novel and powerful network structures for both kinds of MRI reconstruction methods, including CNNs~\cite{Schlemper2017_D5C5,Yang2018_DAGAN}, Recurrent Neural Networks~\cite{Chen2022_Pyramid,Guo2021_Over}, Graph Neural Networks~\cite{Huang2023_VIGU}, recently thriving Transformers~\cite{Feng2021_Task,Huang2022_SwinMR,Huang2022_STGAN,Huang2022_SDAUT,Korkmaz2022_Unsupervised}, etc. 
These rapidly evolving deep learning-based networks enable advances in for MRI reconstruction.

\section*{Methodology}

\begin{figure}[ht]
    \centering
    \includegraphics[width=6in]{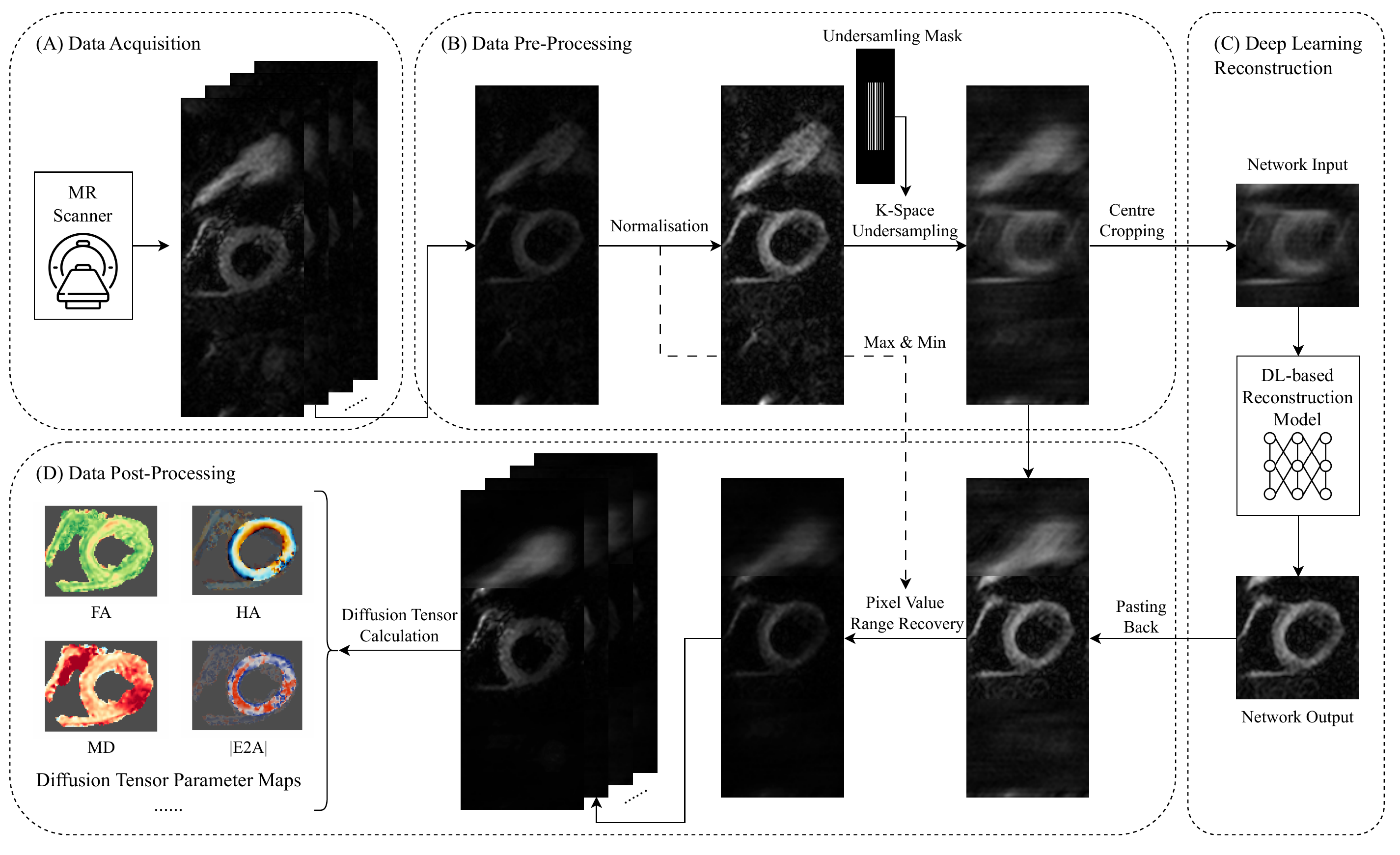}
    \caption{
    The data flow of our implementation for cardiac diffusion tensor imaging data. 
    The whole procedure consists (A) data acquisition, (B) data pre-processing, (C) deep learning-based reconstruction and (D) data post-processing. 
    It is noted that D5C5 does not require the cropping and pasting step and additionally takes the undersampled \textit{k}-space data and the corresponding undersampling mask as input.
    }
    \label{fig:FIG_DATA_FLOW}
\end{figure}

In this study, we implement three deep learning-based MRI reconstruction methods, namely, DAGAN~\cite{Yang2018_DAGAN}, D5C5~\cite{Schlemper2017_D5C5} and SwinMR~\cite{Huang2022_SwinMR}, and assess their performance on cDTI dataset. 
The overall data flow is depicted in Figure~\ref{fig:FIG_DATA_FLOW}.

\subsection*{Data Acquisition}

All data used in this study were approved by the National Research Ethics Service. Written informed consent was obtained from all subjects. 

Retrospectively acquired cDTI data were acquired using Siemens Skyra 3T MRI scanner and Siemens Vida 3T MRI scanner (Siemens AG, Erlangen, Germany). A diffusion-weighted stimulated echo acquisition mode (STEAM) SS-EPI sequence with reduced phase field-of-view and fat saturation. Some MR sequence parameters are listed: 
$\text{TR} = 2 \ \text{RR intervals}$; 
$\text{TE} = 23 \ \text{ms}$; 
SENSE or GRAPPA with $\text{AF} = 2$; 
echo train duration $= 13 \ \text{ms}$; 
spatial resolution $= 2.8 \times 2.8 \times 8.0 \ \text{mm}^\text{3}$.
Diffusion-weighted images were encoded in six directions with diffusion-weighted of $\text{b} = 150 \ \text{and} \ 600 \ \text{sec/mm}^\text{2}$ (namely b150 and b600) in a short-axis mid-ventricular slice. Reference images, namely b0, were also acquired with with a minor diffusion weighting. 

We used 481 cDTI cases including 2 cardiac phases, i.e., diastole ($n = 232$) and systole ($n = 249$), for the experiments section.
The dataset contains 
241 healthy cases, 
31 amyloidosis (AMYLOID) cases, 
47 dilated cardiomyopathy (DCM) cases, 
35 in-recovery DCM (rDCM) cases,
39 hypertrophic cardiomyopathy (HCM) cases,
48 HCM genotype-positive–phenotype-negative (HCM G+P-) cases, 
and 40 acute myocardial infarction (MI) cases.
The overall data distribution of our dataset is shown in Table~\ref{tab:TABEL_DATASET_OVERVIEW}. 
The detailed data distribution per cohort and cardiac phase can be found in Table~\ref{tab:TABEL_DATASET_DETAIL} in Supplementary.

This work separately discussed the reconstruction of systole and diastole cases. For each deep learning-based methods, two network weights were trained for either systole or diastole reconstruction.
In the training stage, we applied 5-fold-cross-validation strategy, using 169 diastole cases (TrainVal-D) or 183 systole cases (TrainVal-S).
In the testing stage, four testing sets were utilised, including mixed ordinary testing set with diastole cases (Test-D) or systole cases (Test-S) and out-of-distribution MI testing set with diastole cases (Test-MI-D) or systole cases (Test-MI-S).
According to Table~\ref{tab:TABEL_DATASET_DETAIL}, Test-D and Test-S includes the data of Health, AMYLOID, rDCM, DCM, HCM and HCM G+P-, which are also included in the TrainVal. 
For further examining the model robustness and ability to handle out-of-the-distribution data, Test-MI dataset includes only MI cases, which are `invisible' for models during the training stage.

\input{TABLE/TABEL_DATASET_OVERVIEW}

\subsection*{Data Pre-Processing}

In the data pre-processing stage, all DWIs (b0, b150 and b600) were processed following the same protocol.

The pixel intensity ranges of DWIs vary considerably across different b-values. 
To address this, We normalised all DWIs in the dataset to a pixel intensity range of $0 \sim 1$ using the max-min method, while the maximum and minimum pixel values of all DWIs were recorded for the pixel intensity range recovery at the beginning of the data post-processing stage.

In our dataset, the majority of DWIs have a resolution of $256 \times 96$, while a small subset of 2D slices exhibit a resolution of $256 \times 88$. 
In order to standardise the resolution, we zero-padded the edges of the images with a resolution of $256 \times 88$ to achieve a resolution of $256 \times 96$.

In this study, GRAPPA-like Cartesian \textit{k}-space undersampling masks with AF $\times 2$, $\times 4$ and $\times 8$, generated by the official protocol of fastMRI dataset~\cite{Zbontar2018_fastMRI}, were applied to simulate the \textit{k}-space undersampling process.
Since all the 2D slices have been reconstructed with zero-padding factor of two, the phase encoding (PE) of our undersampling masks was set to 48 instead of 96, for a more realistic simulation.
The undersampling masks were then zero-padded from $128 \times 48$ to $256 \times 96$ as shown in Figure~\ref{fig:FIG_DATA_FLOW}. More details regarding the undersampling masks can be found in Figure~\ref{fig:FIG_UNDERSAMPLING_MASK} in Supplementary.

For DAGAN and SwinMR, DWIs were further cropped to $96 \times 96$, as both models only support square-shaped input images.

\subsection*{Deep Learning-Based Cardiac Diffusion Tensor Imaging Reconstruction}

In this stage, deep learning-based models were utilised to took the \textit{k}-space undersampled data as the input and produced the reconstructed MR images. We implemented and evaluated three deep learning-based models, namely DAGAN~\cite{Yang2018_DAGAN}, D5C5~\cite{Schlemper2017_D5C5} and SwinMR~\cite{Huang2022_SwinMR} in this stage.

\subsubsection*{DAGAN}

\begin{figure}[ht]
    \centering
    \includegraphics[width=6in]{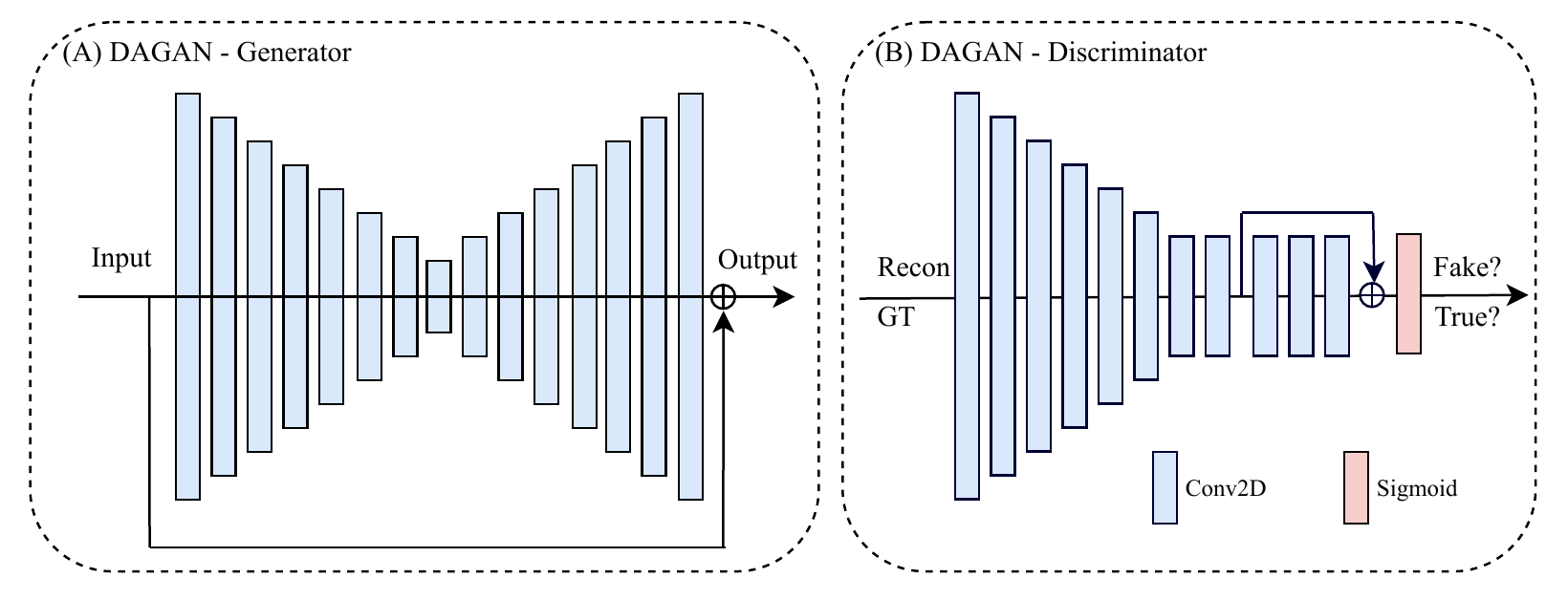}
    \caption{
    The model architecture of DAGAN. 
    (A) the generator of DAGAN is a modified Convolutional Neural Network (CNN)-based U-Net with a residual connection;
    (B) the discriminator of DAGAN is a standard CNN-based classifier.
    Conv2D: 2D convolution layer; Recon: reconstructed MR images; GT: groung truth MR images.
    }
    \label{fig:FIG_MDOEL_STRUCTURE_DAGAN}
\end{figure}

DAGAN~\cite{Yang2018_DAGAN} is a conditional GAN-based and CNN-based model designed for general MRI reconstruction, of which the model structure is presented in Figure.~\ref{fig:FIG_MDOEL_STRUCTURE_DAGAN}. 
DAGAN comprises two components: a generator and a discriminator, which are trained in an adversarial manner as a two-player game.

The generator is a modified CNN-based U-Net~\cite{Ronneberger2015_Unet} with a residual connection~\cite{He2016_Residual}, which takes the \textit{k}-space zero-filled MR images as input and aims to produce reconstructed MR images as close as possible to the ground truth images. 
The discriminator is a standard CNN-based classifier that attempts to distinguish the `fake' reconstructed MR images generated by the generator, from the ground truth MR images.

During the inference stage, only the generator is applied, which takes the ZF MR images as input and outputs the reconstructed images.

DAGAN is trained with a hybrid loss function including an image space $l2$ loss, a frequency space $l2$ loss, a perceptual $l2$ loss based on a pre-trained VGG~\cite{Simonyan2014_VGG}, as well as an adversarial loss~\cite{Goodfellow2020_GAN}. More implement details can be found in the original paper~\cite{Yang2018_DAGAN}.

\subsubsection*{D5C5}

\begin{figure}[ht]
    \centering
    \includegraphics[width=6in]{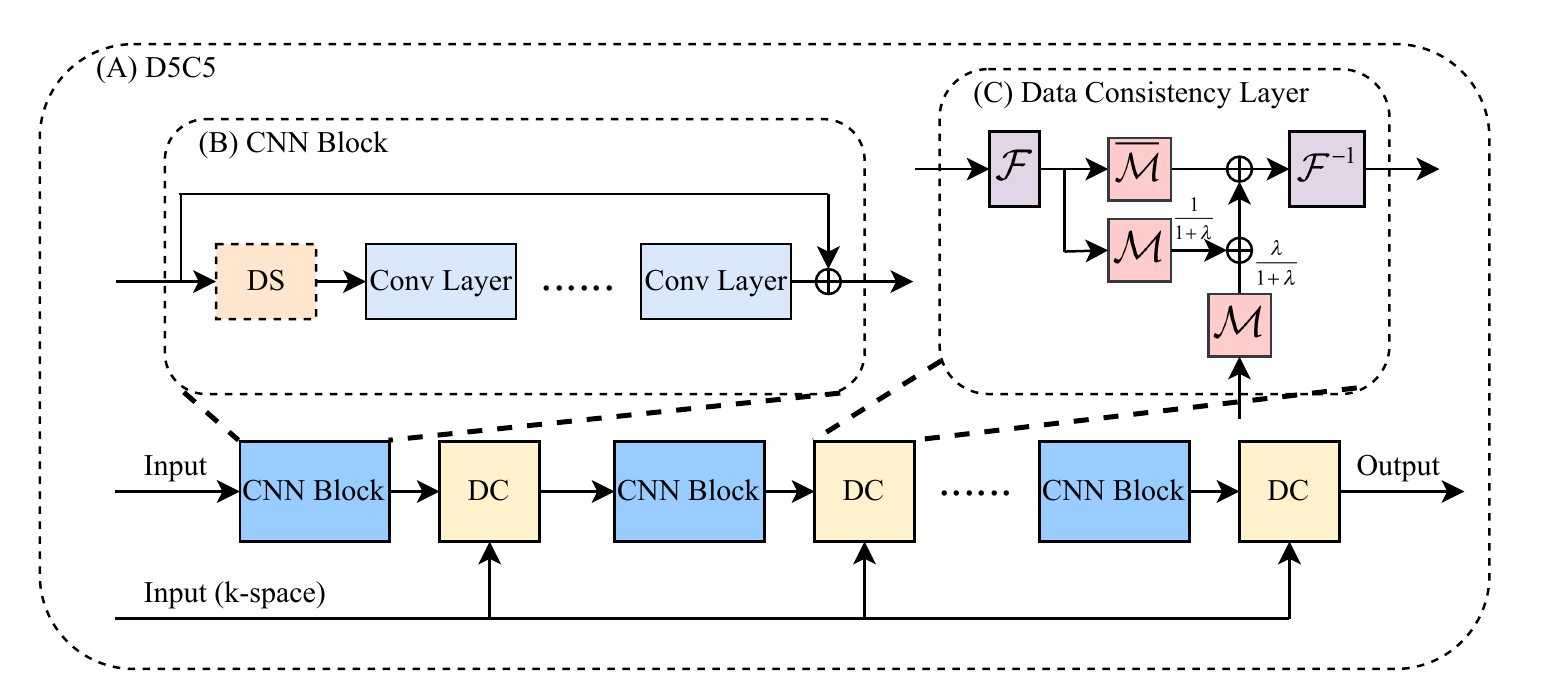}
    \caption{
    (A) The model architecture of D5C5.
    D5C5 has five stages, each comprising a Convolutional Neural Network block (CNN Block) and a data consistency layer (DC).
    (B) The structure of the CNN Block.
    One optional data sharing module (DS) and five convolutional layers (Conv Layers) are included in the CNN Block.
    (C) The structure of the DC. 
    $\mathcal{M}$ denotes the undersampling mask, and $\overline{\mathcal{M}} = \mathcal{I} - \mathcal{M}$.
    $\mathcal{F}$ and $\mathcal{F}^{-1}$ denote the Fourier and inverse Fourier transform.
    $\lambda$ is an adjustable coefficient controlling the level of DC.
    }
    \label{fig:FIG_MDOEL_STRUCTURE_D5C5}
\end{figure}

D5C5~\cite{Schlemper2017_D5C5} is a CNN-based model for MRI reconstruction, with its model structure presented in Figure.~\ref{fig:FIG_MDOEL_STRUCTURE_D5C5}. 
Originally proposed for cine MRI reconstruction, D5C5 also supports general MRI reconstruction.

D5C5 takes the undersampled \textit{k}-space measurement as well as ZF MR images as the input and outputs the reconstructed MR images.
It is composed of multiple stages, each comprising a CNN block and a data consistency (DC) layer.
The CNN block contains a cascade of convolutional layers with Rectifier Linear Units (ReLU) for feature extraction, an optional data sharing (DS) layer for learning spatio-temporal features, as well as a residual connection~\cite{He2016_Residual}. 
The DC layer takes a linear combination between the output of the CNN block and the undersampled \textit{k}-space data, enforcing the consistency between the prediction of CNNs and the original \textit{k}-space measurements.
D5C5 has five stages, with five convolution layers in each CNN block, and no DS layer is applied for our 2D MRI reconstruction task.

D5C5 is trained end-to-end using an image space $l2$ loss function. Further  implementation details can be found in the original paper~\cite{Schlemper2017_D5C5}.

\subsubsection*{SwinMR}

\begin{figure}[ht]
    \centering
    \includegraphics[width=6in]{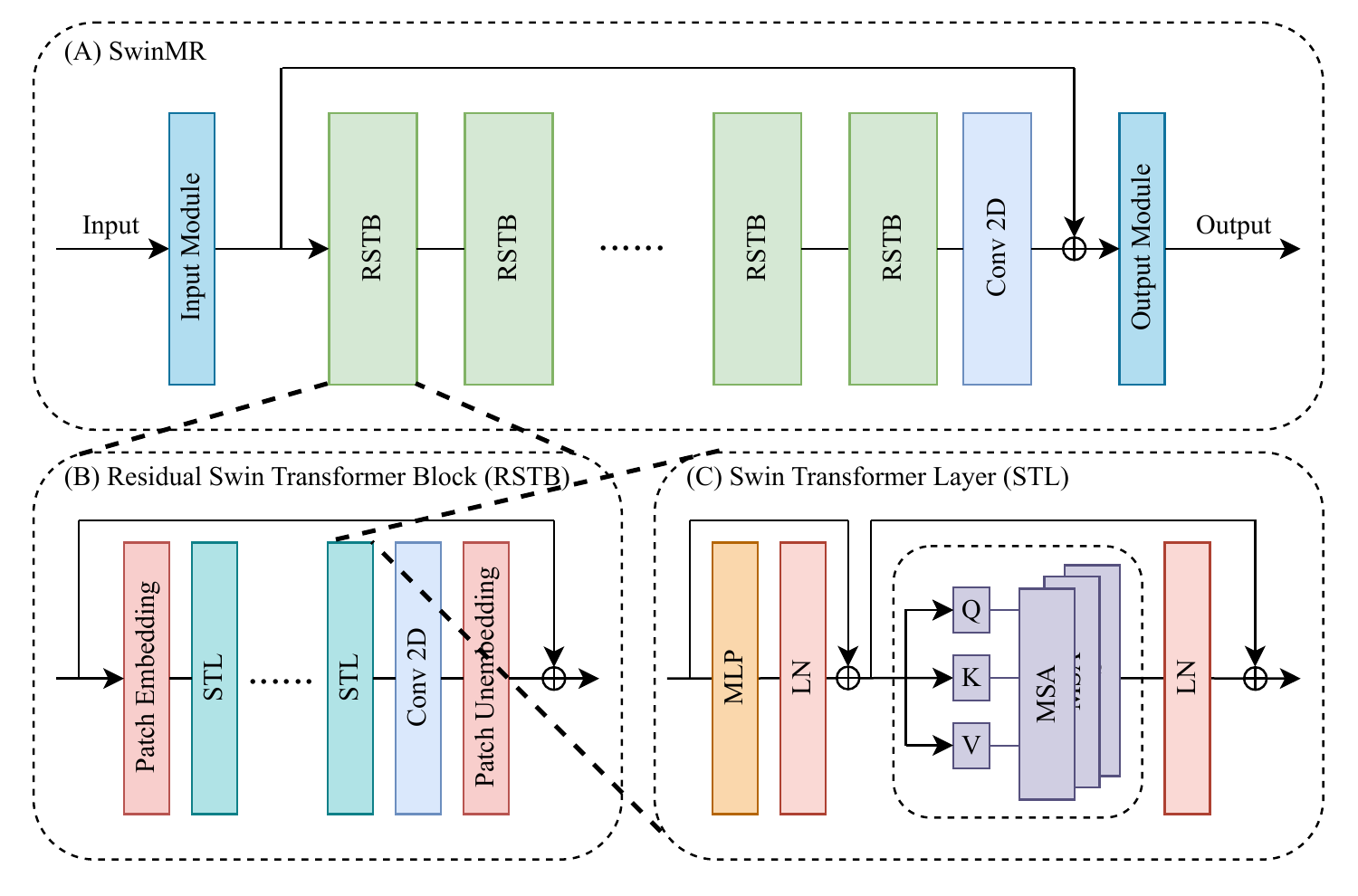}
    \caption{
    (A) The model architecture of SwinMR.
    (B) The structure of the residual Swin Transformer block (RSTB).
    (C) The structure of the Swin Transformer layer (STL).
    Conv2D: 2D convolutional layer.
    MLP: multi-layer perceptron;
    LN: layer normalisation;
    Q: query; K: key; V: value.
    MSA: multi-head self-attention.
    }
    \label{fig:FIG_MDOEL_STRUCTURE_SwinMR}
\end{figure}

SwinMR~\cite{Huang2022_SwinMR} is a Swin Transformer-based model for MRI reconstruction, with its model structure shown in Figure~\ref{fig:FIG_MDOEL_STRUCTURE_SwinMR}. 
SwinMR takes the ZF MR images as the input and directly outputs the reconstructed images.

SwinMR is composed of a CNN-based input module and output module for projecting between the image space and the latent space, a cascade of residual Swin Trasformer blocks (RSTBs), and a convolution layer with a residual connection for feature extraction.
A patch embedding and a patch unembedding layer are placed at the beginning and end of each RSTB, facilitating the inter-conversion of feature maps and sequences, since the computation of Transformers is based on sequences.
Multiple standard Swin Transformer layers (STLs)~\cite{Liu2021_Swin} and a single convolutional layer are applied between the patch embedding and unembedding layer.

SwinMR is trained end-to-end with a hybrid loss function consisting of an image space $l1$ loss, a frequency space $l1$ loss, a perceptual $l1$ loss based on a pre-trained VGG~\cite{Simonyan2014_VGG}. More implementation details can be found in the original paper~\cite{Huang2022_SwinMR}.

\subsection*{Data Post-Processing}

We applied our in-house developed software (MATLAB 2021b, MathWorks, Natick, MA) for cDTI post-processing, following the protocol described in~\cite{Ferreira2014_invivo,Ferreira2022_Accelerating}.
The post-process procedure for reference data includes: 1) manual removal of low-quality DWIs; 2) DWI registration; 3) semi-manual segmentation for left ventricle (LV) myocardium; 4) DT calculation via the LLS fit; 5) DT parameter calculation including FA, MD, HA and E2A. 
The initial post-processing of reference data was performed by either Z.K. (7 years of experience), R.R. (3 years of experience) or M.D. (2 years of experience), and subsequently reviewed by P.F. (10 years of experience). 

For the post-processing of deep leanring-based reconstruction results, the output ($96 \times 96$) of DAGAN and SwinMR were `pasted' back to the corresponding zero-filled images ($256 \times 96$) at their original position. (This process does not affect the final post-processing results since the ROI region is set in the central $96 \times 96$ area.) 

All the DWIs were `anti-normalised' (pixel value range recovery) to their original pixel intensity range using the maximum and minimum values recorded in the pre-processing stage.

The reconstruction results were arranged to construct new reconstruction dataset with the same structure as the reference dataset. The reconstructed dataset was then automatically post-processed following the configuration of reference data (e.g., low-quality removal information, registration shifting, segmentation masks) for a fair comparison.

\section*{Experiments and Results}

In this section, the experimental results are presented from the perspective: 1) the quality of DWIs reconstruction and 2) the quality of DT parameter estimations.

\subsection*{Reconstruction Quality Assessment}

\input{TABLE/TABLE_RECON}

\begin{figure}[t]
    \centering
    \includegraphics[width=6in]{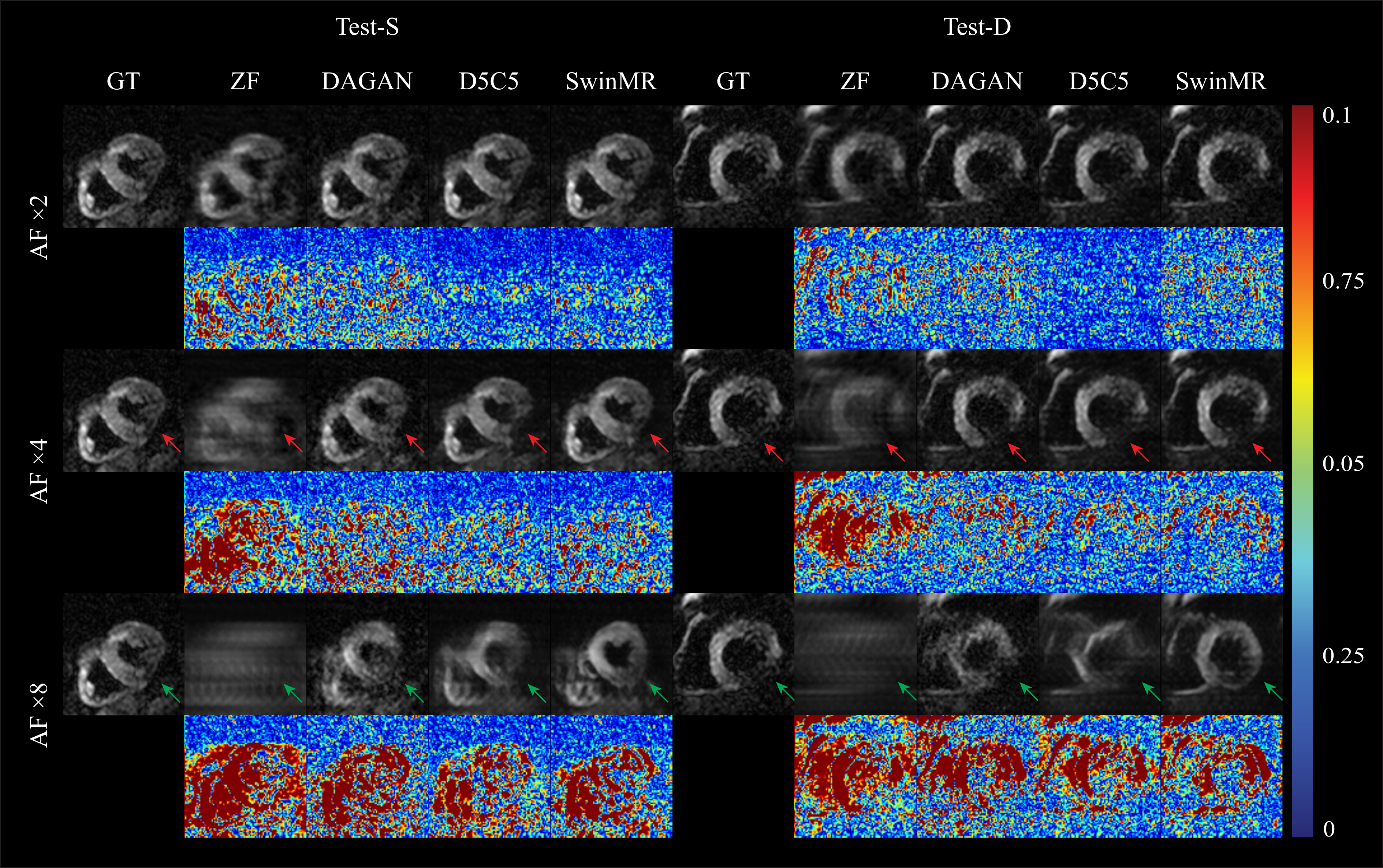}
    \caption{
    The visualised samples of the reconstruction on Test-S and Test-D with the undersampling masks of AF $\times 2$, $\times 4$ and $\times 8$.
    Odd Rows: the ground truth (GT), undersampled \textit{k}-space zero-filled images (ZF) and the reconstruction results of DAGAN, D5C5 and SwinMR.
    Even Rows: the corresponding difference ($\times 10$) of ZF and the reconstruction results of DAGAN, D5C5 and SwinMR.
    Row 1-2: AF $\times 2$;
    Row 3-4: AF $\times 4$;
    Row 5-6: AF $\times 8$;
    Col 1-5: the results on testing set Test-S;
    Col 6-10: the results on testing set Test-D.
    }
    \label{fig:FIG_RECON_VIS}
\end{figure}

In this study, four metrics were considered to assess the reconstruction quality.
Peak Signal-to-Noise Ratio (PSNR) is a simple and commonly used metric for measuring the reconstruction quality, which measures the ratio of the maximum possible power of the signal to the power. 
Higher PSNR value indicates a better reconstruction quality.
Structural Similarity Index (SSIM) is a perceptual-based metric that measures the similarity between two images by comparing their structural information. 
Higher SSIM value indicates a better reconstruction quality.
Learned Perceptual Image Patch Similarity (LPIPS)~\cite{Zhang2018_LPIPS} is a learned metric that measures the perceptual similarity between two images by computing the distance in the latent space using a pre-trained deep neural network.
LPIPS has shown a high correlation with human perceptual judgements of the image similarity.
Lower LPIPS value indicates a better generated images quality.
Fr\'echet Inception Distance (FID)~\cite{Heusel2017_FID} is a learned metric that measures the similarity between two sets of images by comparing their feature statistics, using a pre-trained deep neural network. 
FID has also shown to have high correlation with human perceptual experience.
Lower FID value indicates a better generated images quality.

Quantitative reconstruction results on the Test-S and Test-D datasets are presented in Table~\ref{tab:TABLE_RECON}.
The two-sample t-test was applied for the statistical analysis, and $^{\star}$ in Table~\ref{tab:TABLE_RECON} indicates the specific result distribution is significantly different ($p < 0.05$) from the \textbf{best result} distribution.
Among the evaluated models, D5C5 demonstrates superior fidelity in the reconstruction, while SwinMR provides results with higher perceptual score. 

Visualised samples of the reconstruction results on Test-S and Test-D datasets are shown in Figure~\ref{fig:FIG_RECON_VIS}.

\subsection*{Diffusion Tensor Parameter Quality Assessment}

\input{TABLE/TABLE_DT_SYSTOLE}

We further evaluated the quality of DT parameters, including FA, MD, E2A and HA, after post-processing. 

Differences in DT parameter global mean values between the reference and reconstruction, on systole testing sets (Test-S and Test-MI-S) and diastole testing sets (Test-D and Test-MI-D) are presented in Table~\ref{tab:TABLE_DT_SYSTOLE} and Table~\ref{tab:TABLE_DT_DIASTOLE}, respectively. 
The mean absolute error for FA, MD and the mean absolute angular error for the HA gradient (HA Slope) and E2A were employed to quantify the difference.
The Mann-Whitney test was utilised for the statistical analysis, and $^{\star}$ in Table~\ref{tab:TABLE_DT_SYSTOLE} and Table~\ref{tab:TABLE_DT_DIASTOLE} indicates that the specific error distribution is significantly different ($p < 0.05$) from the \textbf{best results} distribution.
Data point with \colorbox{bk}{green background} indicates that the specific distribution of corresponding DT parameter global mean values is NOT significantly different ($p > 0.05$) from the reference distribution according to the Mann-Whitney Test.

Overall, SwinMR can achieve better or comparable (not significantly different) MD, HA slope and E2A results on all testing sets.
DAGAN can achieve better or comparable (not significantly different) FA results on all testing sets.
D5C5 has provided better results only on Test-S at AF $\times 2$, but it is not significantly better than SwinMR (on MD, HA Slope and E2A) and DAGAN (on FA).

\begin{figure}[t]
    \centering
    \includegraphics[width=6.5in]{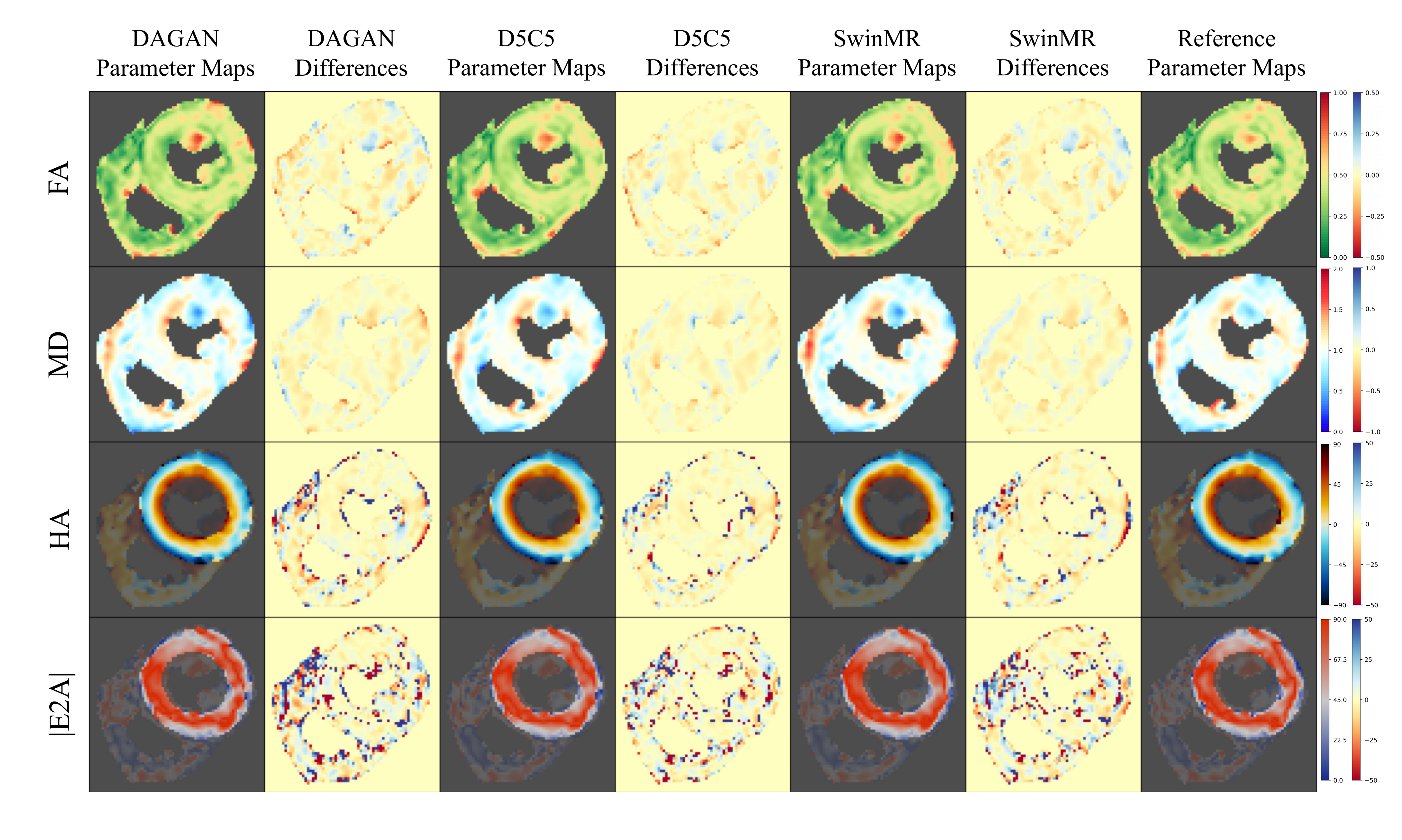}
    \caption{
    Diffusion parameter maps of the reconstruction results (AF $\times 2$) and the reference of a healthy systole case from testing set Test-S.
    Row 1: fractional anisotropy (FA); 
    Row 2: mean diffusivity (MD); 
    Row 3: helix angle (HA); 
    Row 4: absolute value of the second eigenvector (|E2A|). 
    }
    \label{fig:FIG_DT_VIS_AF2_AS_HEALTHY}
\end{figure}

\begin{figure}[t]
    \centering
    \includegraphics[width=6.5in]{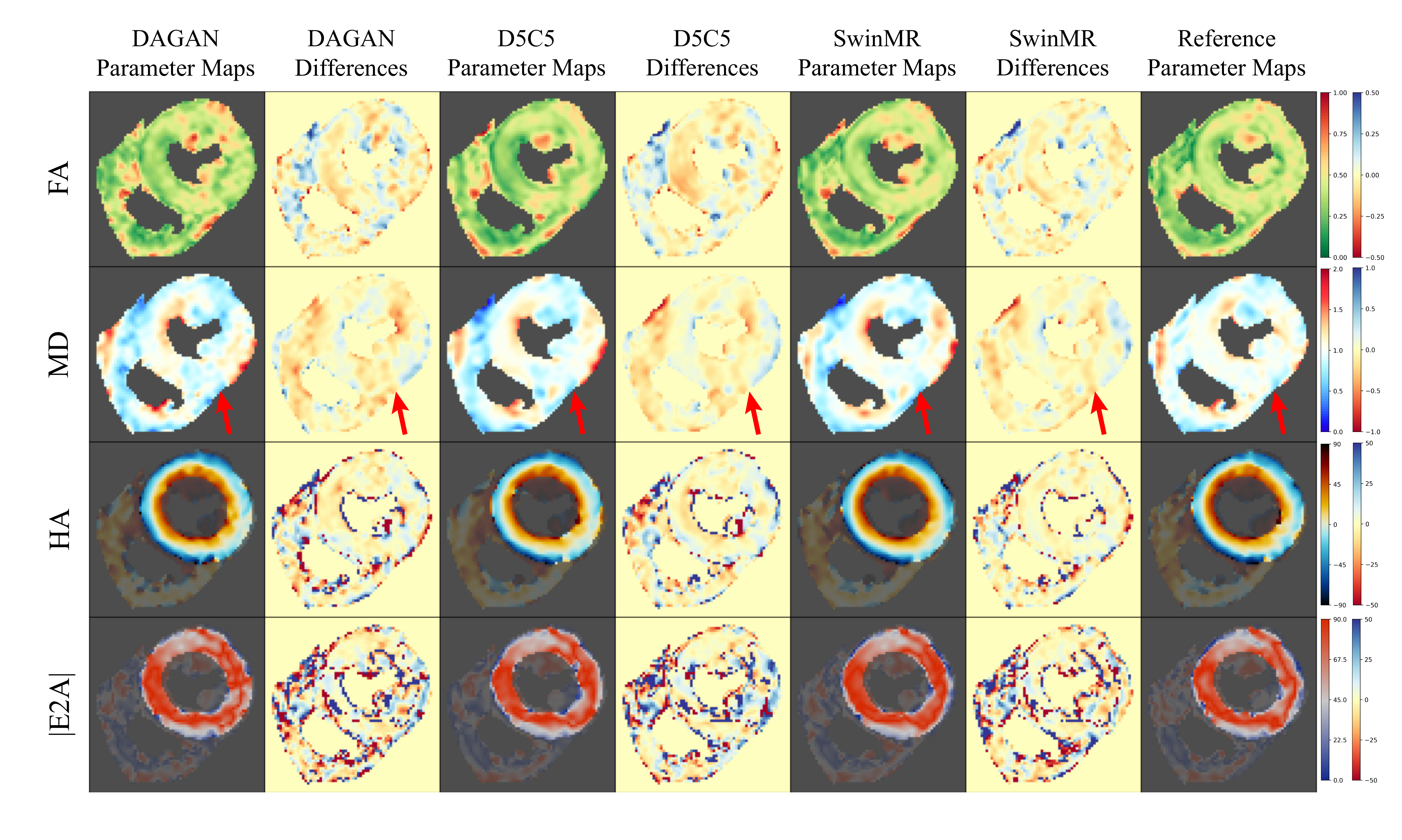}
    \caption{
    Diffusion parameter maps of the reconstruction results (AF $\times 4$) and the reference of a healthy systole case from testing set Test-S.
    Row 1: fractional anisotropy (FA); 
    Row 2: mean diffusivity (MD); 
    Row 3: helix angle (HA); 
    Row 4: absolute value of the second eigenvector (|E2A|).
    }
    \label{fig:FIG_DT_VIS_AF4_AS_HEALTHY}
\end{figure}

\begin{figure}[t]
    \centering
    \includegraphics[width=6.5in]{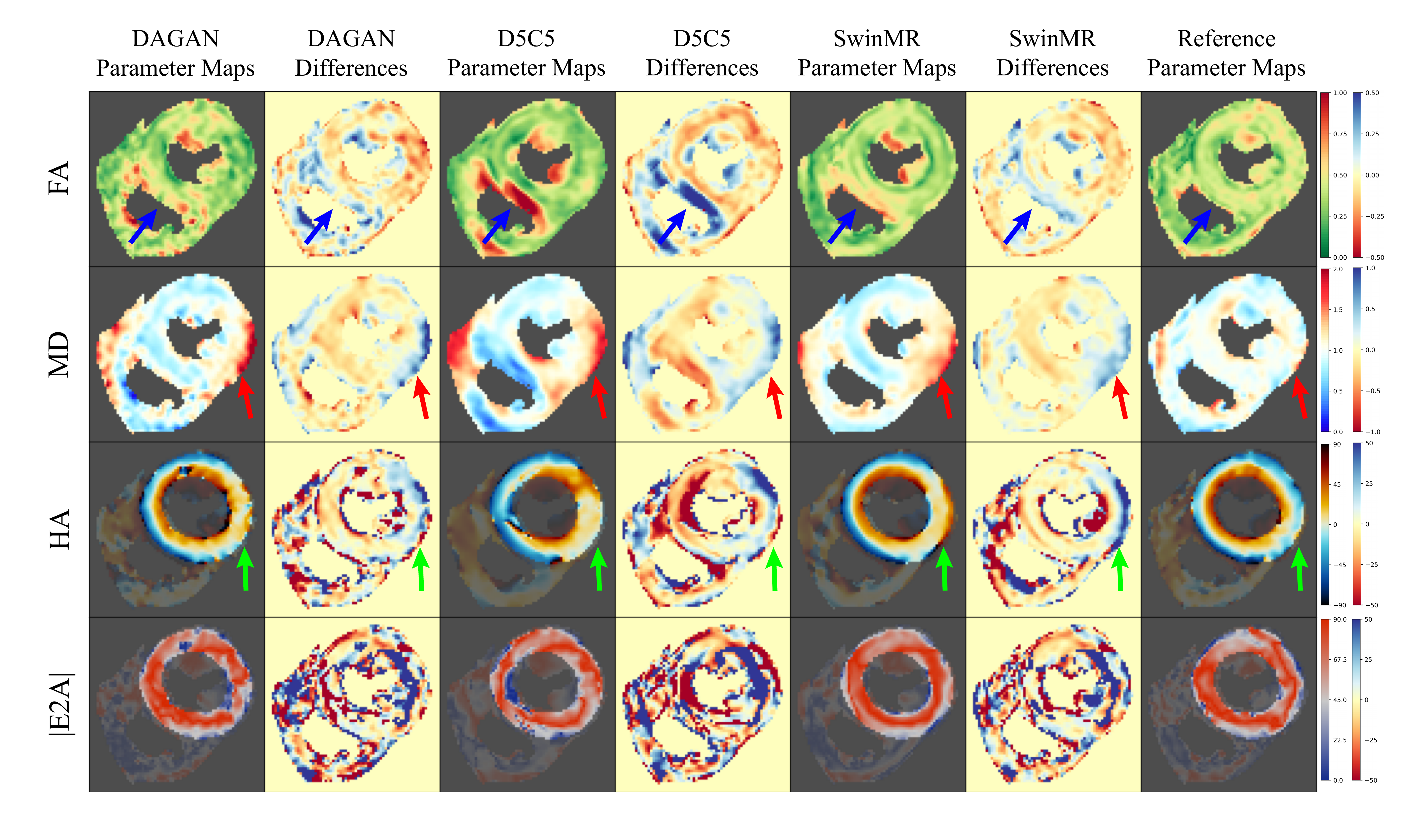}
    \caption{
    Diffusion parameter maps of the reconstruction results (AF $\times 8$) and the reference of a healthy systole case from testing set Test-S.
    Row 1: fractional anisotropy (FA); 
    Row 2: mean diffusivity (MD); 
    Row 3: helix angle (HA); 
    Row 4: absolute value of the second eigenvector (|E2A|). 
    }
    \label{fig:FIG_DT_VIS_AF8_AS_HEALTHY}
\end{figure}

Some cases of visualised DT parameter maps are presented in this study, including FA, MD, HA and absolute value of E2A (|E2A|).
The DT parameter maps of a systole healthy case from Test-S with different AFs are shown in 
Figure~\ref{fig:FIG_DT_VIS_AF2_AS_HEALTHY} (AF$\times 2$), 
Figure~\ref{fig:FIG_DT_VIS_AF4_AS_HEALTHY} (AF$\times 4$),
and
Figure~\ref{fig:FIG_DT_VIS_AF8_AS_HEALTHY} (AF$\times 8$).
The DT parameter maps of a diastole healthy case from Test-D with different AFs are shown in 
Figure~\ref{fig:FIG_DT_VIS_AF2_AD_HEALTHY} (AF$\times 2$), 
Figure~\ref{fig:FIG_DT_VIS_AF4_AD_HEALTHY} (AF$\times 4$),
and
Figure~\ref{fig:FIG_DT_VIS_AF8_AD_HEALTHY} (AF$\times 8$).
The DT parameter maps of a systole MI case from Test-MI-S with different AFs are shown in 
Figure~\ref{fig:FIG_DT_VIS_AF2_AS_MI} (AF$\times 2$), 
Figure~\ref{fig:FIG_DT_VIS_AF4_AS_MI} (AF$\times 4$),
and
Figure~\ref{fig:FIG_DT_VIS_AF8_AS_MI} (AF$\times 8$).
The DT parameter maps of a diastole MI case from Test-MI-D with different AFs are shown in 
Figure~\ref{fig:FIG_DT_VIS_AF2_AD_MI} (AF$\times 2$), 
Figure~\ref{fig:FIG_DT_VIS_AF4_AD_MI} (AF$\times 4$),
and
Figure~\ref{fig:FIG_DT_VIS_AF8_AD_MI} (AF$\times 8$).

\section*{Discussion}

In this study, we have investigated the performance of deep learning-based methods in the context of cDTI reconstruction.
We have implemented three deep learning-based MRI reconstruction methods, namely DAGAN, D5C5 and SwinMR, on our cDTI dataset.
Experimental results have been reported from the perspective of reconstruction quality assessment and DT estimation quality assessments.

According to Table~\ref{tab:TABLE_RECON}, 
for the reconstruction tasks with undersampling masks of AF $\times 2$ and AF $\times 4$, D5C5 has achieved superior PSNR and SSIM, while SwinMR has achieved better deep learning-based perceptual scores, i.e., LPIPS and FID.
For the reconstruction tasks at AF $\times 8$, SwinMR has outperformed other methods across all the metrics applied.

According to Figure~\ref{fig:FIG_RECON_VIS},
for the reconstruction task of AF $\times 2$, all three methods have produced fairly good visual reconstruction results. 
For the reconstruction task of AF $\times 4$, all three methods have successfully recovered overall structure information, whereas they have behaved differently in the recovery of the high-uncertainty area. For example in the experiment on Test-S at AF $\times 4$ (Row 3-4, Col 1-5, Figure~\ref{fig:FIG_RECON_VIS}), the red arrows indicate the high-uncertainty area on the LV myocardium due to the signal loss. DAGAN has provided a noisy estimation while SwinMR has clearly preserved this part of information. However, the results of D5C5 have missed the information in this area.
For reconstruction task of AF $\times 8$, neither of three methods is able to produce visually satisfied reconstruction results. 
For example in the experiment on Test-S with AF $\times 8$ (Row 5-6, Col 1-5, Figure~\ref{fig:FIG_RECON_VIS}), a large amount of visible aliasing artefacts along the PE direction has remained in the reconstruction results of both D5C5 and SwinMR, with D5C5 performing relatively worse than SwinMR. DAGAN, to some extent, has eliminated the aliasing artefacts at the expense of the increased noise, leading to a low-SNR reconstruction.
Regarding the recovery of high-uncertainty area, both DAGAN and D5C5 have failed to preserve the information in this area. SwinMR can retain most information of this area, but meanwhile it has produced `fake' estimation (green arrow).

For the fidelity of the reconstruction, D5C5 has shown superiority on the condition of a relative lower AF, whereas this superiority has been observed disappearance on the condition of a relative higher AF. 
This phenomenon is caused by the utilisation of DC module in D5C5 (Figure~\ref{fig:FIG_MDOEL_STRUCTURE_D5C5}), which combines the \textit{k}-space measurements information with the CNN estimation to keep the consistency. 
According to Figure~\ref{fig:FIG_UNDERSAMPLING_MASK}, in the reconstruction task at a relative lower AF, a large proportion of information in the final output of D5C5 is provided by the DC module, whereas this proportion is significantly decreased in a relative higher AF reconstruction task (AF $\times 8$).
Therefore, this kind of unrolling-based methods with DC module is more suitable for the reconstruction at a relative lower AF.

For the perceptual score of the reconstructions, experiments have shown SwinMR outperforms D5C5 and DAGAN on metrics LPIPS and FID. 
However, even though the perceptual score has a high correlation with the observation of human, it is not always equivalent to a better reconstruction quality~\cite{Huang2022_STGAN}. According to Figure~\ref{fig:FIG_RECON_VIS} (green arrow), SwinMR has learnt to estimate a `fake' reconstruction detail for a higher perceptual score, which is totally unacceptable and dangerous for clinical use.
We believes this phenomenon is caused by the nature of the Transformer applied in SwinMR, which is powerful enough to estimate and generate details that does not exist originally. In addition, the utilisation of the perceptual VGG-based loss restricts SwinMR to produce more \emph{perceptual-similar} reconstruction instead of \emph{pixel-wise-similar} reconstruction.

In general, the differences for tensor parameter global mean values between reference and reconstruction results tend to increase as the AF rises. 
Concerning the global mean values of FA,
DAGAN has demonstrated superiority on the Test-S and Test-MI-S, with its superiority growing as the AF increases.
On the Test-D and Test-MI-D, the three methods have yielded similar results, with no statistically significant difference observed.
Regarding the global mean values of MD, 
D5C5 and SwinMR have outperformed DAGAN across all the testing sets.
Specifically, D5C5 has delivered better results on Test-S, while SwinMR has excelled on Test-MI-S.
On the Test-D and Test-MI-D, SwinMR and D5C5 have achieved similar results with no statistical difference at AF $\times 2$ and AF $\times 4$, while SwinMR has surpassed D5C5 at a higher AF (AF $\times 8$).
For the global mean values of HA Slope,
it is clear that SwinMR has outperformed DAGAN and D5C5 on all testing sets, 
with its superiority being statistically significant on Test-S and Test-D.
In terms of the global mean values of E2A,
generally SwinMR has achieved better or comparable results among the three methods, but the differences are typically not statistically significant.

Generally, the quality of DT parameter maps have decreased as the AF increases. 
We believes that at AF $\times 2$ and AF $\times 4$, the DT parameter maps calculated by these three methods can achieve similar level with the reference. 
For the MI cases from out-of-the-distribution testing set Test-MI, these three methods can successfully preserve the information in lesion area for clinical use.
For example at AF $\times 2$, all three methods have provided visually similar DT parameters maps with the reference (Figure~\ref{fig:FIG_DT_VIS_AF2_AS_HEALTHY} and Figure~\ref{fig:FIG_DT_VIS_AF2_AS_MI}). 
At AF $\times 4$, all three methods can recover most information of the DT parameters maps. 
DAGAN tends to produce noisier DT parameter maps, while SwinMR and D5C5 tend to produce the smoother DT parameter maps, which matches the results from the reconstruction quality assessment.
We can observed from the MD map and its corresponding error map, that the vertical aliasing (along PE) direction has affected the DT parameter maps (Figure~\ref{fig:FIG_DT_VIS_AF4_AS_HEALTHY}, red arrows).
The intensity of the MI area in the MD map of DAGAN has had a trend to decrease, while D5C5 and SwinMR has clearly preserved it (Figure~\ref{fig:FIG_DT_VIS_AF4_AS_MI}, red arrows).

However, at AF $\times 8$, the quality of DT parameter maps have significantly gone worse, which also matches the results from the reconstruction quality assessment.
For the FA map, a band of higher FA is expected to be observed in the mesocardium for a healthy heart~\cite{McGill2015_Heterogeneity}. 
However, DAGAN and D5C5 have failed to recover the band of higher FA, where DAGAN has produced very noisy FA map, and D5C5 over-smoothed the FA map and wrongly estimate a highlight area (Figure~\ref{fig:FIG_DT_VIS_AF8_AS_MI}, blue arrows).
For the MD map, the affect from the aliasing that has been observed at AF $\times 4$, has gone more severe.
In the healthy case, the highlight area has wrongly appeared in MD maps from all three methods, which is unacceptable for clinical use and may lead to misdiagnosis (Figure~\ref{fig:FIG_DT_VIS_AF8_AS_HEALTHY}, red arrows).
In the MI case, the MI lesion area tends to decrease for all the methods, especially for the results of DAGAN, where the lesion area has nearly disappeared (Figure~\ref{fig:FIG_DT_VIS_AF8_AS_MI}, red arrows).
For the HA map, it has been observed that SwinMR can produce relatively smooth HA map, while DAGAN can only reconstruct very noisy one. 
However, the direction of HA has been wrong estimated in the epicardium of the healthy case (Figure~\ref{fig:FIG_DT_VIS_AF8_AS_HEALTHY}, green arrows). This is not acceptable for clinical use and easier to lead to misdiagnosis such as MI.
For the |E2A| map, DAGAN tends to reconstruct a noisy map, while SwinMR tends to produce a smooth map. 
All three maps can reconstruct similar results with reference even at AF $\times 8$.

Through our experiments, we have demonstrated that the models discussed in this paper can be effectively applied for clinical use at AF $\times 2$ and AF $\times 4$. However, at AF $\times 8$, the performance of these three models, including the best-performing SwinMR, has still remained limited.

We hope that this study will serve as a baseline for the future cDTI reconstruction model development. 
Our findings have indicated that there are still limitations when directly applying these general MRI reconstruction methods for cDTI reconstruction.

\emph{There is an absence of restrictions on diffusion.} 
The loss functions utilised in the three models discussed in this study all rely on image domain loss, with D5C5 and DAGAN additionally incorporating the frequency domain loss and the perceptual loss. 
In other words, these is no diffusion information restriction implemented during the model training stage. 
For further work, the diffusion tensor or parameter maps can be jointly considered into the loss function. Moreover, physical constraints on diffusion can be also incorporated into the training stage.

\emph{There is a trade-off between perceptual performance and quantitative performance.}
Cardiac diffusion tensor MRI is a quantitative technique, which places greater emphasis on the accuracy contrast, pixel intensity range, and pixel-wise fidelity, also referred to as pixel-wise distance.
However, the models discussed in this study were originally designed for structural MRI, and they tend to pay more attention on the `perceptual-similarity', which can be regarded as the latent space distance.
A trade-off exists between pixel-wise fidelity and perceptual-similarity~\cite{Blau2018_Perception}. For example, blurred images generally exhibit better pixel-wise fidelity, while the images with clear but `fake' details tend to have better perceptual-similarity~\cite{Huang2022_STGAN}. 
Such `fake' details can sometimes be harmful for clinical use.
Consequently, for further work, more efforts should be made to consider about how to improve the pixel-wise fidelity rather than the perceptual-similarity, or how to prevent the appearance of the `fake' information.

\emph{There is a gap between current DT evaluation methods and the true quality of cDTI reconstruction.} 
This study has revealed that the global mean value of diffusion parameters is not always accurate or sensitive enough to evaluate the diffusion tensor quality. For example, Table~\ref{tab:TABLE_DT_SYSTOLE} indicates no statistically significant difference in MD between reconstruction results (even including ZF) and the reference on Test-S, whereas the Figure~\ref{fig:FIG_DT_VIS_AF8_AS_HEALTHY} shows that the MD maps are entirely unacceptable. This discrepancy arises because the MD value increases and decreases in different parts of the MD map, while the global mean value maintains relative consistency, rendering the global mean MD ineffective in reflecting the quality of the final DT estimation.
For future work, apart from the visualised assessment, we will applied the down-stream task assessment, e.g., utilising a pre-trained pathology classification or detection mdoel to evaluate the reconstruction quality. Theoretically, better classification or detection accuracy corresponds to improved reconstruction results.

There are still limitations for this study. 
1) The size of testing sets is not sufficiently large. 
The relative small testing sets enlarge the randomness of experimental results and reduce the reliability of statistical tests.
In future studies, we will expand our dataset, and provide more accurate results.
2) Our simulation experiment is based on the retrospective \textit{k}-space undersampling on single-channel DWIs that have been reconstructed by the MR scanner. 
The retrospective undersampling step itself has removed a large amount of noise, leading to unrealistic post-processing results.
In future studies, we will conduct our experiment on prospectively-acquired \textit{k}-space raw data.

\section*{Conclusion}

In conclusion, we have investigated the application of deep learning-based methods for accelerating cDTI reconstruction, which has significant potential for improving the integration of cDTI into routine clinical practice. 
Our study focuses on three different models, namely D5C5, DAGAN, and SwinMR, which have been evaluated on cDTI datasets with the AF of $\times 2$, $\times 4$, and $\times 8$. 
The results have demonstrated that the examined models can be effectively utilised for clinical use at AF $\times 2$ and AF $\times 4$, with SwinMR being the recommended optimal approach. 
However, at AF $\times 8$, the performance of all models has remained limited, and further research is required to improve their performance at a relative higher AF.

\bibliography{reference}

\section*{Acknowledgements (not compulsory)}

This study was supported in part by the UKRI Future Leaders Fellowship (MR/V023799/1), BHF (RG/19/1/34160), the ERC IMI (101005122), the H2020 (952172), the MRC (MC/PC/21013), the Royal Society (IEC/NSFC/211235), the NVIDIA Academic Hardware Grant Program, EPSRC (EP/V029428/1, EP/S026045/1, EP/T003553/1, EP/N014588/1, EP/T017961/1), and the Cambridge Mathematics of Information in Healthcare Hub (CMIH) Partnership Fund.

\section*{Author contributions statement}

Must include all authors, identified by initials, for example:
J.H. and G.Y. conceptualised the study, P.F.F, A.A.R., C.B.S., A.D.S.,S.N.V., and G.Y. guided the methodology development, J.H., P.F.F., S.N.V., and G.Y. conceived the experiments, J.H., P.F.F., L.W., Y.W. and G.Y. conducted the experiments, J.H., P.F.F., L.W., Y.W., S.N.V., and G.Y. analysed the results, Z.K., M.D., R.R., R.D.S., and D.J.P. provided the data including labelling and clinical guidance. All authors reviewed the manuscript. 

\section*{Additional information}
\textbf{Competing interests} D.J.P receives research support from Siemens and is a stockholder and director of Cardiovascular Imaging Solutions. The RBH CMR group receives research support from Siemens Healthineers. Others have no Conflicts of Interest.



\clearpage


\noindent \textbf{\raggedright\sffamily\bfseries\fontsize{20}{25}\selectfont 
Supplemental Materials of \\
Deep Learning-based Diffusion Tensor Cardiac Magnetic Resonance Reconstruction: A Comparison Studies
\\
}%

\setcounter{equation}{0}
\setcounter{figure}{0}
\setcounter{table}{0}
\makeatletter
\renewcommand{\theequation}{S\arabic{equation}}
\renewcommand{\thefigure}{S\arabic{figure}}
\renewcommand{\thetable}{S\arabic{table}}


\vspace{5em}

\section*{Methodology}

\begin{figure}[ht]
    \centering
    \includegraphics[width=4in]{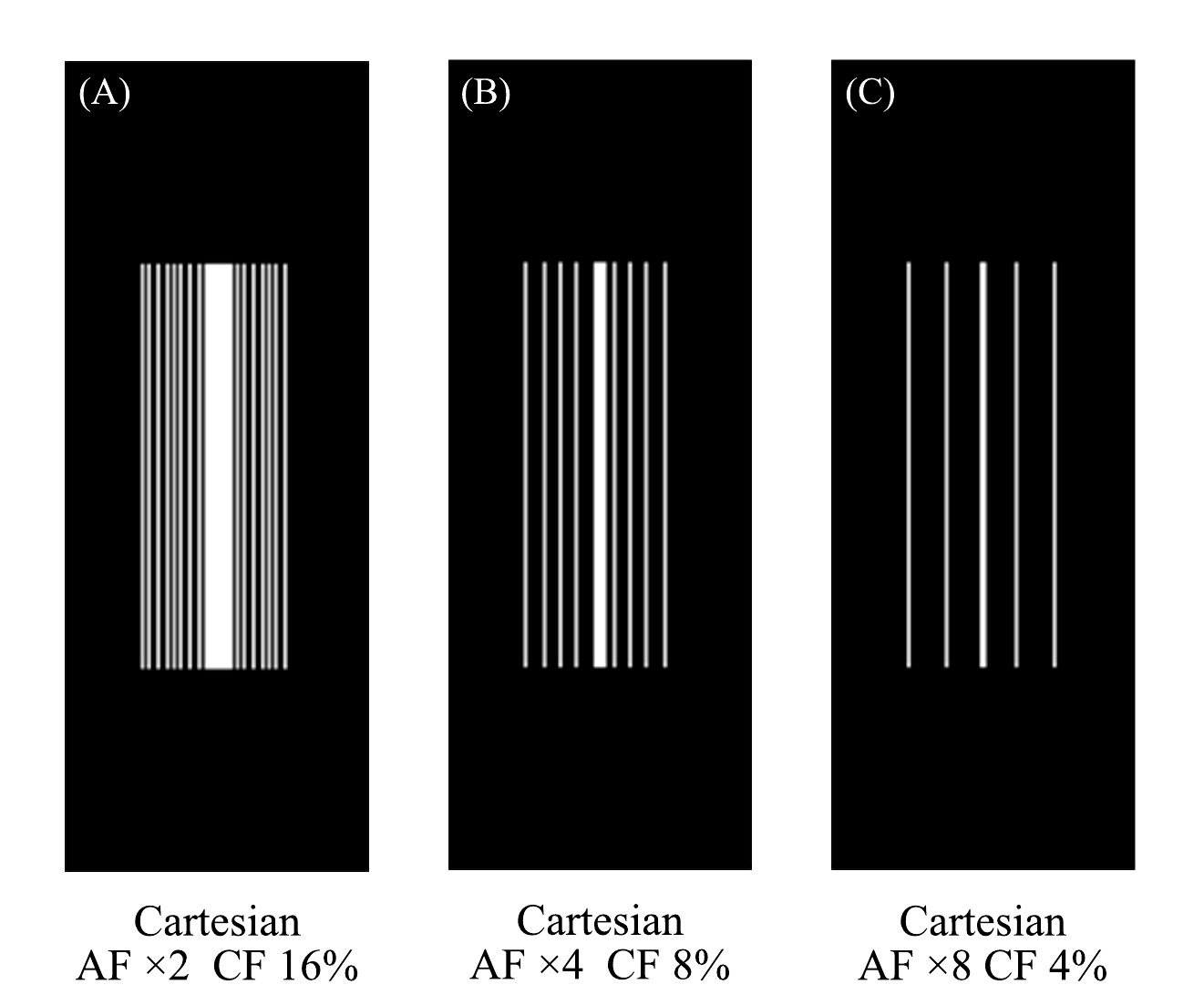}
    \caption{
    Three Cartesian \textit{k}-space undersampling masks applied in this work. 
    AF: acceleration factor; 
    CF: centre factor.
    }
    \label{fig:FIG_UNDERSAMPLING_MASK}
\end{figure}

\input{TABLE/TABEL_DATASET_DETAIL}

\clearpage

\section*{Experiments and Results}

\input{TABLE/TABLE_DT_DIASTOLE}

\begin{figure}[t]
    \centering
    \includegraphics[width=6.5in]{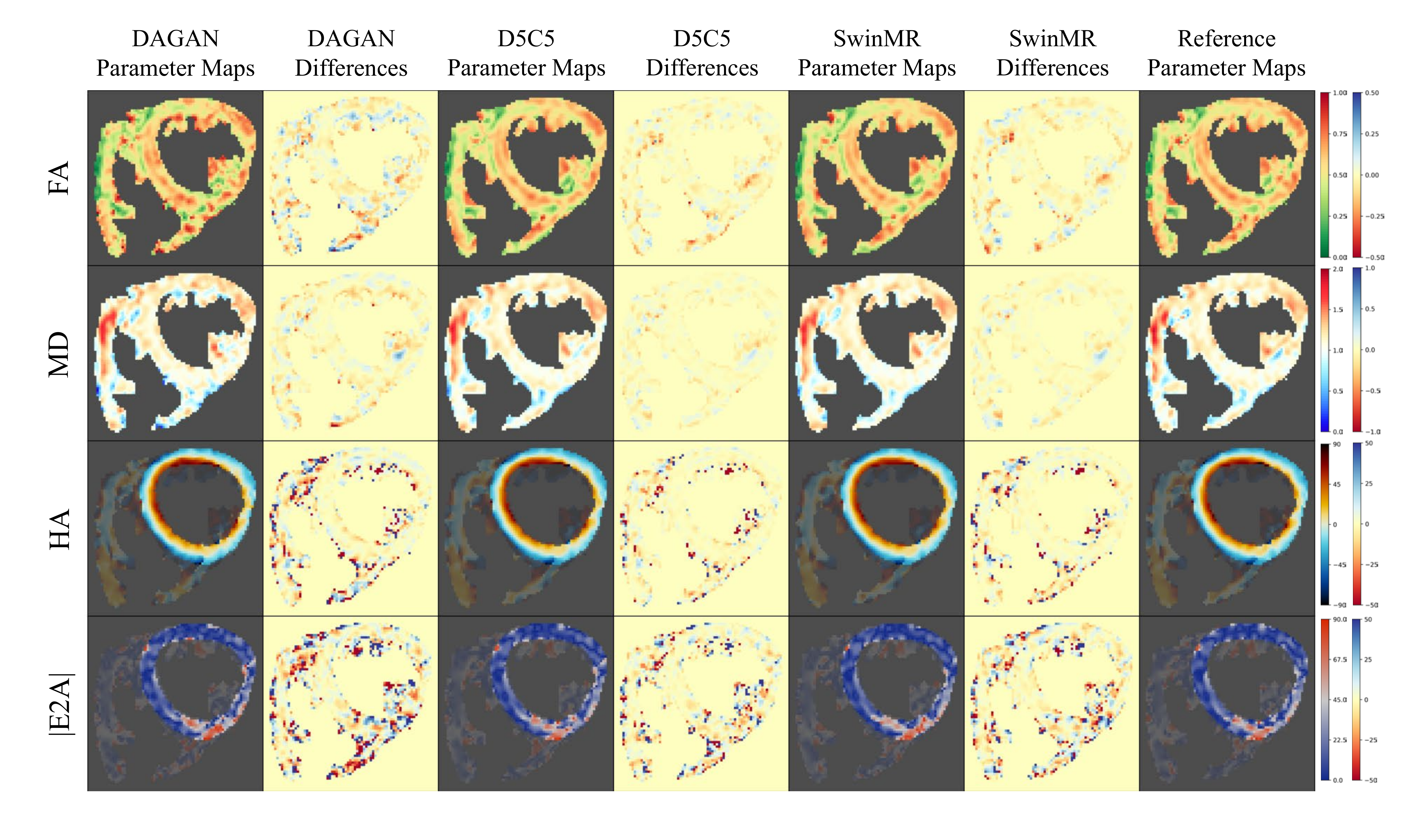}
    \caption{
    Diffusion parameter maps of the reconstruction results (AF $\times 2$) and the reference of a healthy diastole case from testing set Test-D.
    Row 1: fractional anisotropy (FA); 
    Row 2: mean diffusivity (MD); 
    Row 3: helix angle (HA); 
    Row 4: absolute value of the second eigenvector (|E2A|).
    }
    \label{fig:FIG_DT_VIS_AF2_AD_HEALTHY}
\end{figure}

\begin{figure}[t]
    \centering
    \includegraphics[width=6.5in]{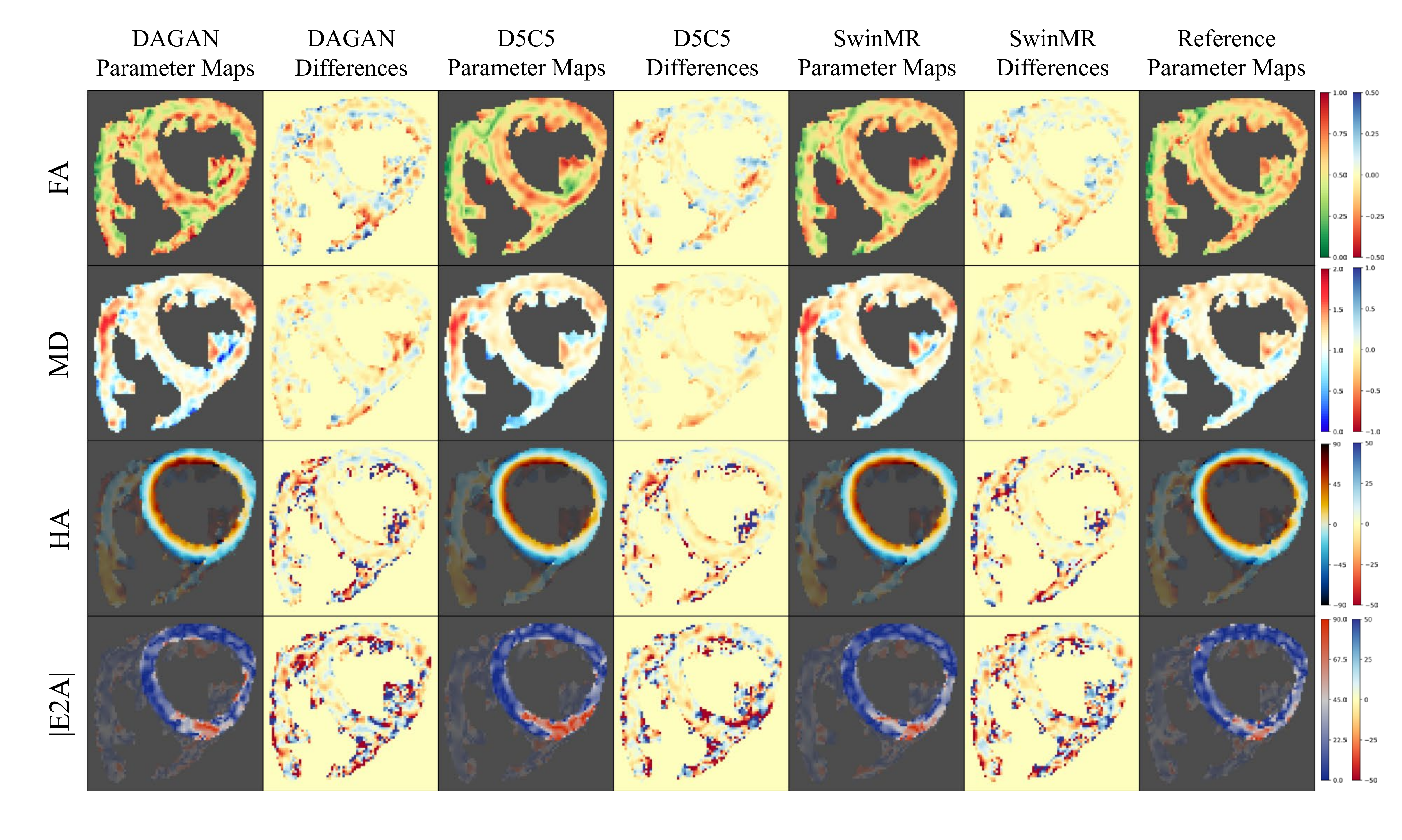}
    \caption{
    Diffusion parameter maps of the reconstruction results (AF $\times 4$) and the reference of a healthy diastole case from testing set Test-D.
    Row 1: fractional anisotropy (FA); 
    Row 2: mean diffusivity (MD); 
    Row 3: helix angle (HA); 
    Row 4: absolute value of the second eigenvector (|E2A|).
    }
    \label{fig:FIG_DT_VIS_AF4_AD_HEALTHY}
\end{figure}

\begin{figure}[t]
    \centering
    \includegraphics[width=6.5in]{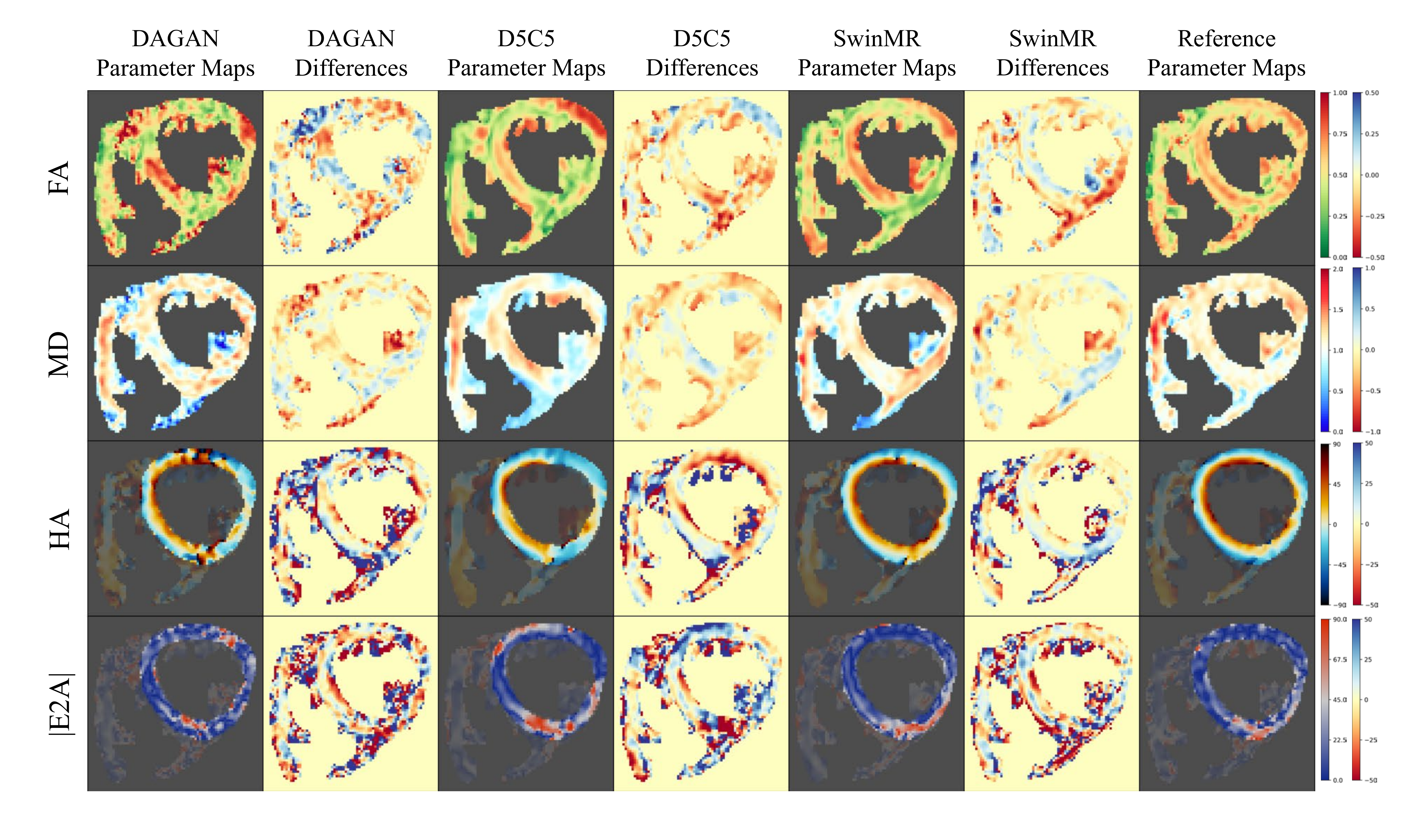}
    \caption{
    Diffusion parameter maps of the reconstruction results (AF $\times 8$) and the reference of a healthy diastole case from testing set Test-D.
    Row 1: fractional anisotropy (FA); 
    Row 2: mean diffusivity (MD); 
    Row 3: helix angle (HA); 
    Row 4: absolute value of the second eigenvector (|E2A|).
    }
    \label{fig:FIG_DT_VIS_AF8_AD_HEALTHY}
\end{figure}

\begin{figure}[t]
    \centering
    \includegraphics[width=6.5in]{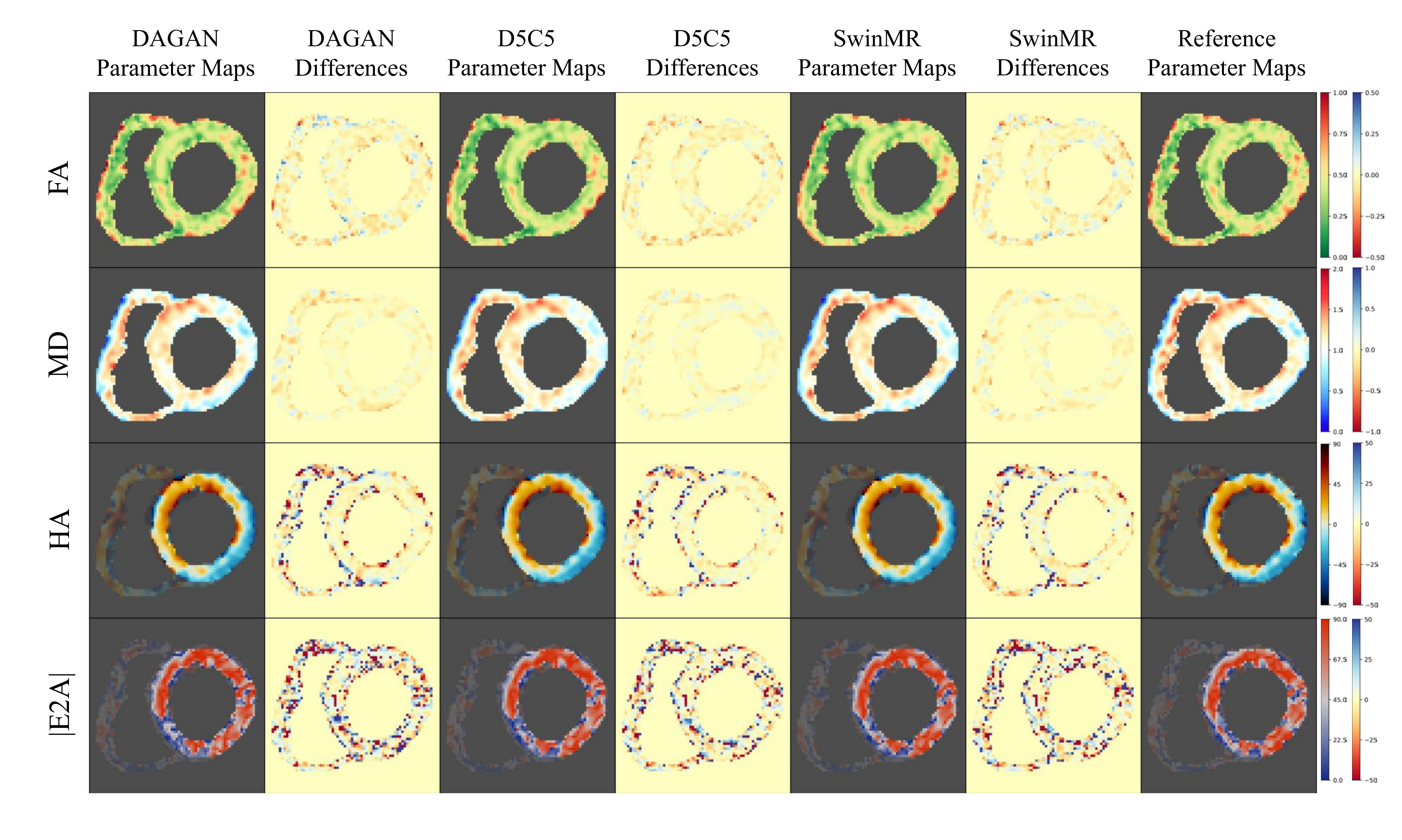}
    \caption{
    Diffusion parameter maps of the reconstruction results (AF $\times 2$) and the reference of a acute myocardial infarction (MI) systole case from testing set Test-MI-S.
    Row 1: fractional anisotropy (FA); 
    Row 2: mean diffusivity (MD); 
    Row 3: helix angle (HA); 
    Row 4: absolute value of the second eigenvector (|E2A|).
    }
    \label{fig:FIG_DT_VIS_AF2_AS_MI}
\end{figure}

\begin{figure}[t]
    \centering
    \includegraphics[width=6.5in]{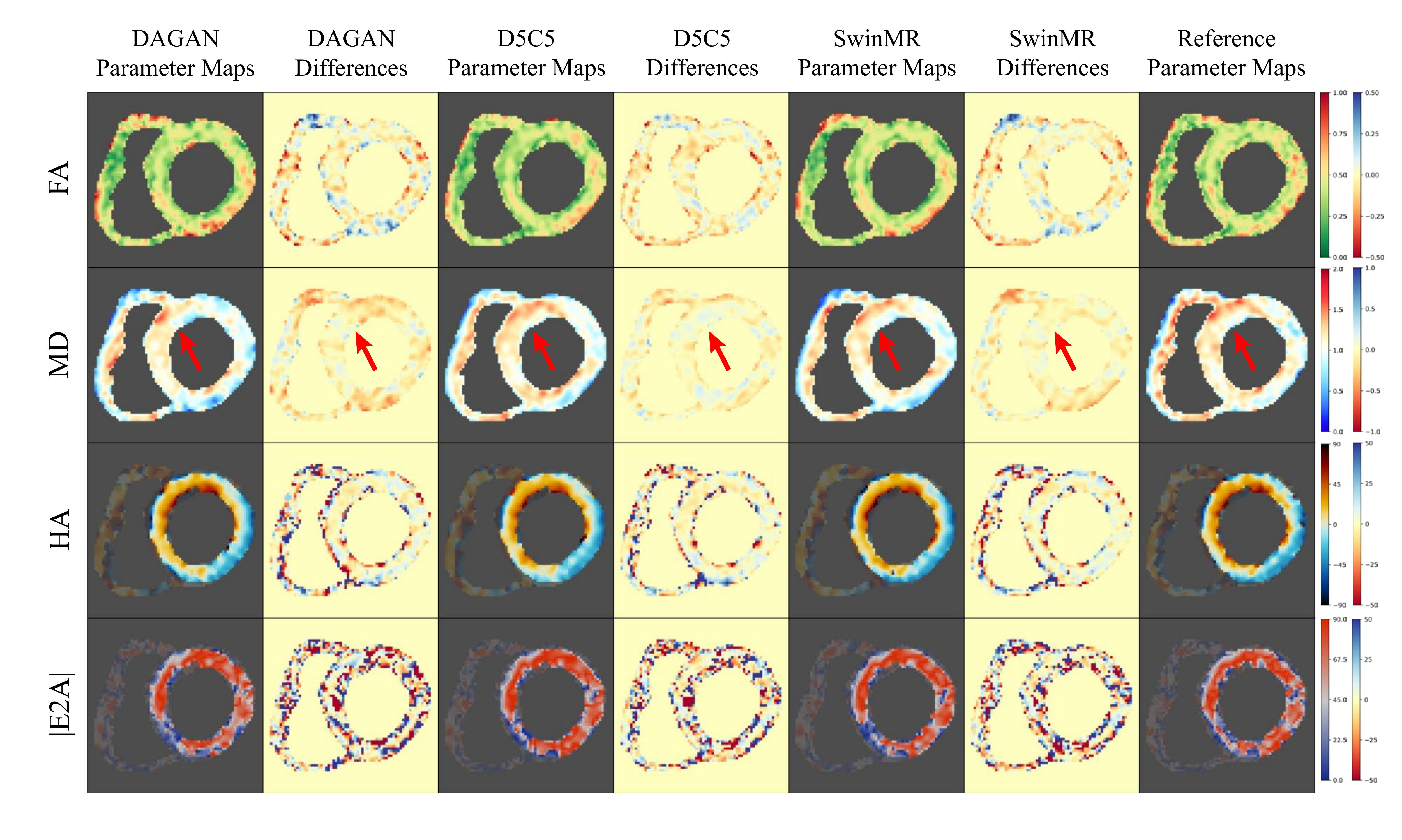}
    \caption{
    Diffusion parameter maps of the reconstruction results (AF $\times 4$) and the reference of a acute myocardial infarction (MI) systole case from testing set Test-MI-S.
    Row 1: fractional anisotropy (FA); 
    Row 2: mean diffusivity (MD); 
    Row 3: helix angle (HA); 
    Row 4: absolute value of the second eigenvector (|E2A|). 
    }
    \label{fig:FIG_DT_VIS_AF4_AS_MI}
\end{figure}

\begin{figure}[t]
    \centering
    \includegraphics[width=6.5in]{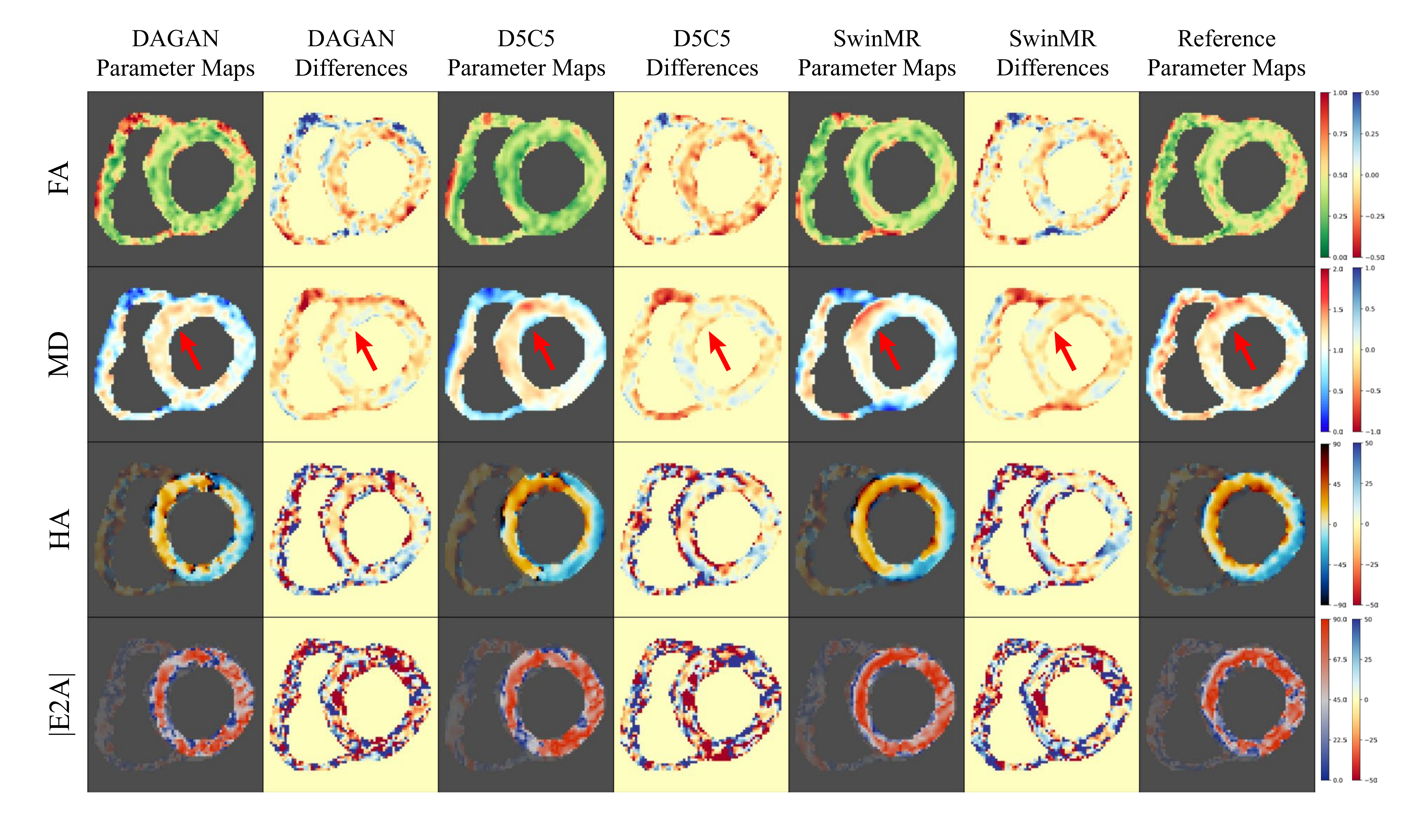}
    \caption{
    Diffusion parameter maps of the reconstruction results (AF $\times 8$) and the reference of a acute myocardial infarction (MI) systole case from testing set Test-MI-S.
    Row 1: fractional anisotropy (FA); 
    Row 2: mean diffusivity (MD); 
    Row 3: helix angle (HA); 
    Row 4: absolute value of the second eigenvector (|E2A|). 
    }
    \label{fig:FIG_DT_VIS_AF8_AS_MI}
\end{figure}

\begin{figure}[t]
    \centering
    \includegraphics[width=6.5in]{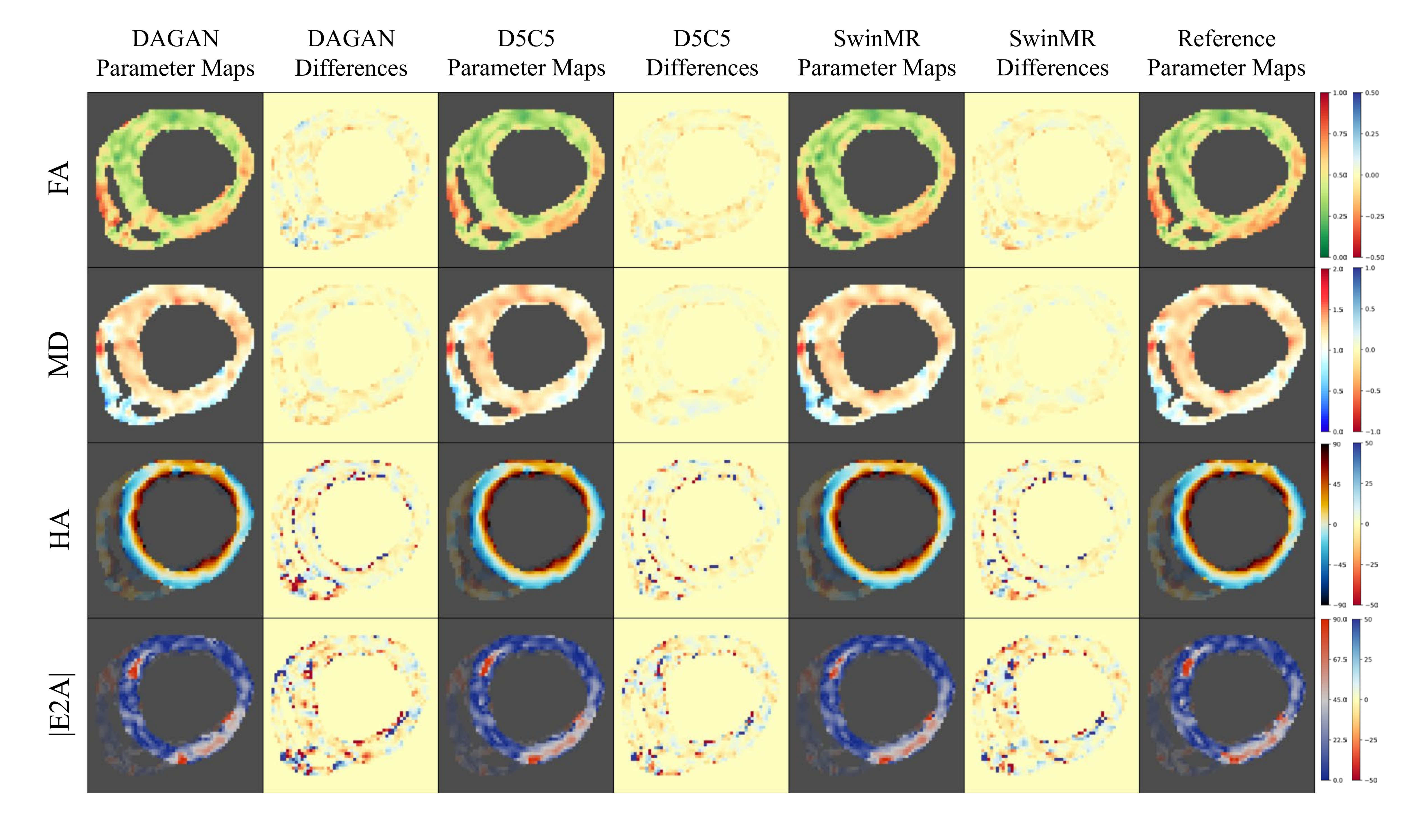}
    \caption{
    Diffusion parameter maps of the reconstruction results (AF $\times 2$) and the reference of a acute myocardial infarction (MI) diastole case from testing set Test-MI-D.
    Row 1: fractional anisotropy (FA); 
    Row 2: mean diffusivity (MD); 
    Row 3: helix angle (HA); 
    Row 4: absolute value of the second eigenvector (|E2A|). 
    }
    \label{fig:FIG_DT_VIS_AF2_AD_MI}
\end{figure}

\begin{figure}[t]
    \centering
    \includegraphics[width=6.5in]{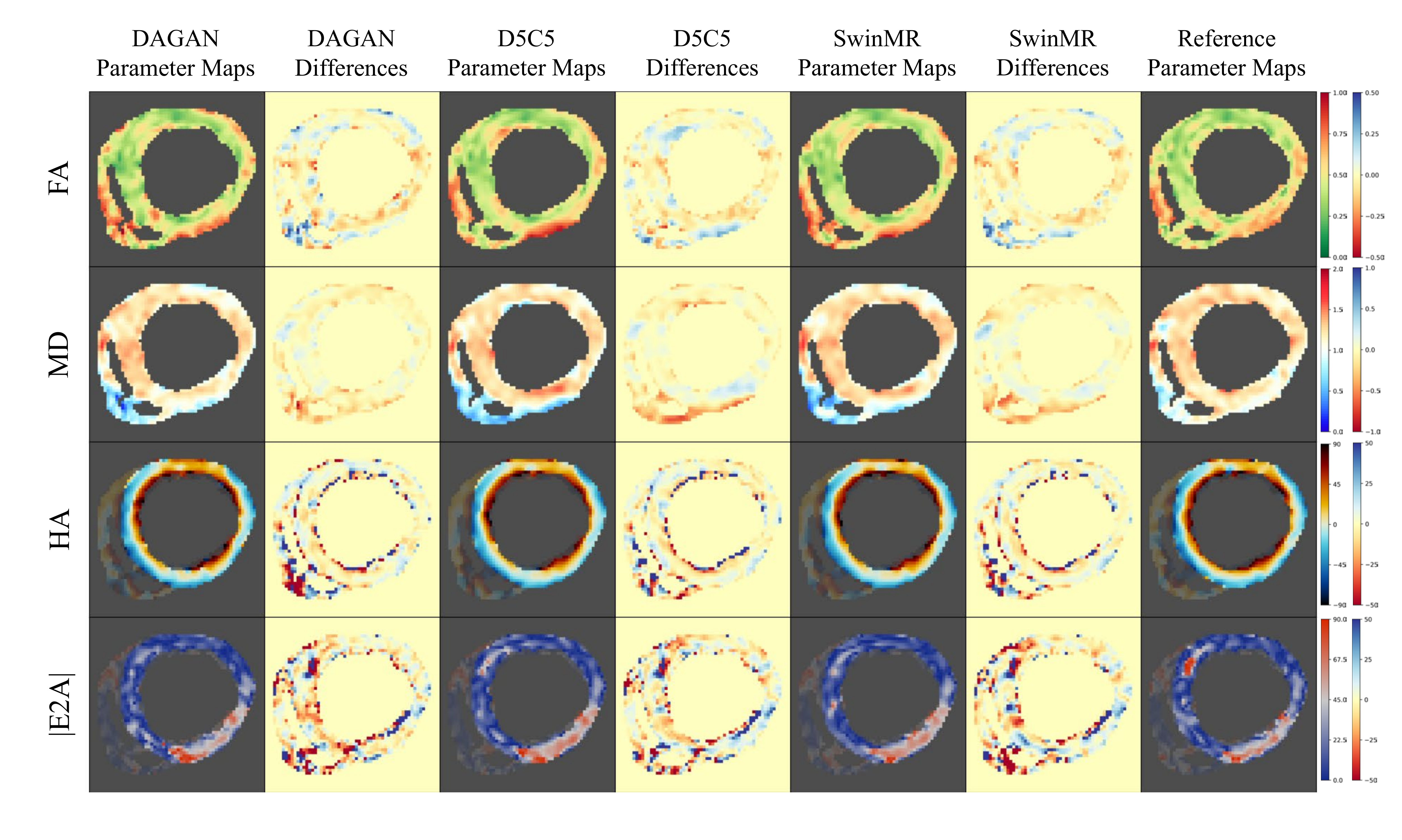}
    \caption{
    Diffusion parameter maps of the reconstruction results (AF $\times 4$) and the reference of a acute myocardial infarction (MI) diastole case from testing set Test-MI-D.
    Row 1: fractional anisotropy (FA); 
    Row 2: mean diffusivity (MD); 
    Row 3: helix angle (HA);
    Row 4: absolute value of the second eigenvector (|E2A|).
    }
    \label{fig:FIG_DT_VIS_AF4_AD_MI}
\end{figure}

\begin{figure}[t]
    \centering
    \includegraphics[width=6.5in]{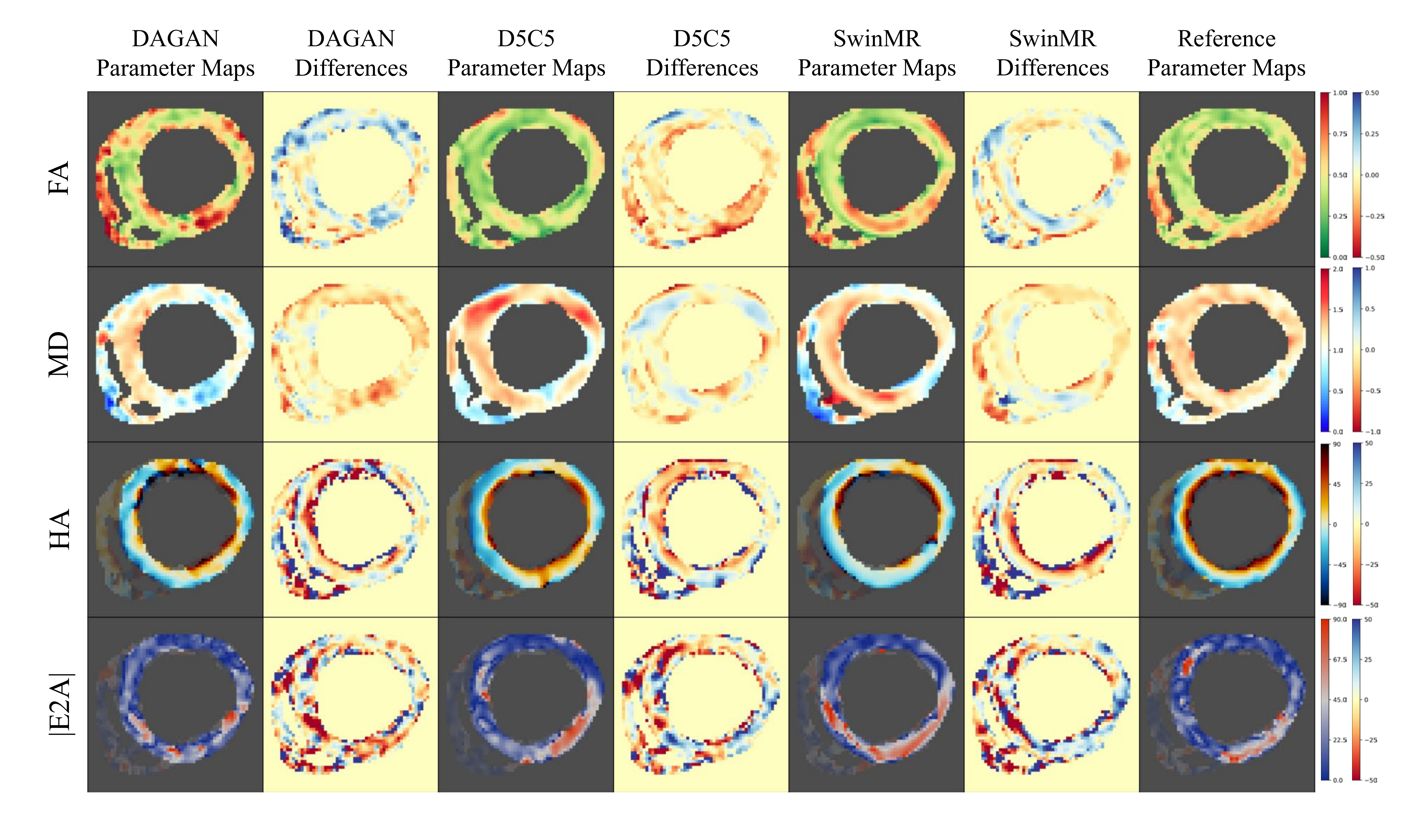}
    \caption{
    Diffusion parameter maps of the reconstruction results (AF $\times 8$) and the reference of a acute myocardial infarction (MI) diastole case from testing set Test-MI-D.
    Row 1: fractional anisotropy (FA); 
    Row 2: mean diffusivity (MD); 
    Row 3: helix angle (HA); 
    Row 4: absolute value of the second eigenvector (|E2A|).
    }
    \label{fig:FIG_DT_VIS_AF8_AD_MI}
\end{figure}

\end{document}

%% file: TABLE/TABEL_DATASET_OVERVIEW.tex
\begin{table}[htbp]
  \centering
  \caption{The overview of the dataset.}
    \setlength{\tabcolsep}{5.5mm}{
    \begin{tabular}{ccccccccc}
    \toprule
    \multirow{2}[4]{*}{Cardaic Phase} & \multicolumn{2}{c}{Total} & \multicolumn{2}{c}{TrainVal} & \multicolumn{2}{c}{Test} & \multicolumn{2}{c}{Test-MI} \\
\cmidrule{2-9}          & Case  & Slice & Case  & Slice & Case  & Slice & Case  & Slice \\
    \midrule
    Diastole & 272   & 22509 & 169   & 14182 & 43    & 3630  & 20    & 1470 \\
    Systole & 290   & 23654 & 183   & 15054 & 46    & 3938  & 20    & 1470 \\
    \bottomrule
    \end{tabular}%
    }%
  \label{tab:TABEL_DATASET_OVERVIEW}%
\end{table}%

%% file: TABLE/TABLE_RECON.tex
\begin{table}[htbp]
  \centering
  \caption{
  The quantitative reconstruction results on the testing sets Test-S and Test-D with undersampling masks of the acceleration factor (AF) $\times 2$, $\times 4$ and $\times 8$.
  SSIM, PSNR and LPIPS results are quoted as `mean (standard deviation)'.
  $^{\star}$ indicates the specific distribution is significantly different ($p < 0.05$) from the \textbf{best results} distribution by the two-sample t-test. 
  }
    \resizebox{\textwidth}{!}{%
    \begin{tabular}{c|cccc|cccc}
    \toprule
    \textbf{Metrics} & \multicolumn{4}{c}{\textbf{Test-S}} & \multicolumn{4}{c}{\textbf{Test-D}} \\
    \midrule
    \textbf{AF x2} & ZF    & DAGAN & D5C5  & SwinMR & ZF    & DAGAN & D5C5  & SwinMR \\
    \midrule
    SSIM $\uparrow$  & 0.819 (0.042) $^{\star}$ & 0.857 (0.026) $^{\star}$ & \textbf{0.931 (0.030)} & 0.919 (0.031) $^{\star}$ & 0.819 (0.044) $^{\star}$ & 0.851 (0.025) $^{\star}$ & \textbf{0.932 (0.031)} & 0.921 (0.034) $^{\star}$ \\
    PSNR $\uparrow$  & 25.76 (2.27) $^{\star}$ & 27.86 (1.59) $^{\star}$ & \textbf{31.80 (2.45)} & 30.83 (2.50) $^{\star}$ & 26.41 (2.38) $^{\star}$ & 28.25 (1.52) $^{\star}$ & \textbf{32.43 (2.70)} & 31.61 (2.80) $^{\star}$ \\
    LPIPS $\downarrow$ & 0.149 (0.037) $^{\star}$ & 0.060 (0.024) $^{\star}$ & \textbf{0.050 (0.026)} & \textbf{0.050 (0.023)} & 0.148 (0.034) $^{\star}$ & 0.066 (0.028) $^{\star}$ & 0.053 (0.032) & \textbf{0.052 (0.029)} \\
    FID $\downarrow$   & 82.29 & 33.1  & 18.53 & \textbf{17.2} & 101.55 & 38.84 & 25.83 & \textbf{22.7} \\
    \midrule
    \textbf{AF x4} & ZF    & DAGAN & D5C5  & SwinMR & ZF    & DAGAN & D5C5  & SwinMR \\
    \midrule
    SSIM $\uparrow$  & 0.663 (0.068) $^{\star}$ & 0.751 (0.039) $^{\star}$ & \textbf{0.849 (0.044)} & 0.842 (0.048) $^{\star}$ & 0.668 (0.065) $^{\star}$ & 0.783 (0.039) $^{\star}$ & \textbf{0.860 (0.047)} & 0.851 (0.051) $^{\star}$ \\
    PSNR $\uparrow$  & 20.92 (2.34) $^{\star}$ & 24.20 (1.62) $^{\star}$ & \textbf{26.85 (2.14)} & 26.56 (2.19) $^{\star}$ & 21.86 (2.40) $^{\star}$ & 25.66 (1.76) $^{\star}$ & \textbf{28.27 (2.37)} & 27.75 (2.51) $^{\star}$ \\
    LPIPS $\downarrow$ & 0.321 (0.040) $^{\star}$ & 0.117 (0.041) $^{\star}$ & 0.092 (0.030) $^{\star}$ & \textbf{0.090 (0.030)} & 0.313 (0.041) $^{\star}$ & 0.090 (0.032) & 0.091 (0.039) $^{\star}$ & \textbf{0.089 (0.039)} \\
    FID $\downarrow$   & 218.18 & 61.9  & 38.42 & \textbf{29.48} & 212.61 & 51.41 & 44.64 & \textbf{36.17} \\
    \midrule
    \textbf{AF x8} & ZF    & DAGAN & D5C5  & SwinMR & ZF    & DAGAN & D5C5  & SwinMR \\
    \midrule
    SSIM $\uparrow$  & 0.529 (0.084) $^{\star}$ & 0.579 (0.054) $^{\star}$ & 0.680 (0.067) $^{\star}$ & \textbf{0.719 (0.074)} & 0.544 (0.080) $^{\star}$ & 0.595 (0.054) $^{\star}$ & 0.689 (0.064) $^{\star}$ & \textbf{0.720 (0.072)} \\
    PSNR $\uparrow$  & 18.26 (2.37) $^{\star}$ & 20.15 (1.73) $^{\star}$ & 21.52 (2.09) $^{\star}$ & \textbf{22.32 (2.19)} & 19.33 (2.40) $^{\star}$ & 20.99 (1.80) $^{\star}$ & 22.63 (2.20) $^{\star}$ & \textbf{23.17 (2.30)} \\
    LPIPS $\downarrow$ & 0.491 (0.040) $^{\star}$ & 0.212 (0.059) $^{\star}$ & 0.197 (0.049) $^{\star}$ & \textbf{0.165 (0.046)} & 0.473 (0.037) $^{\star}$ & 0.199 (0.061) $^{\star}$ & 0.196 (0.055) $^{\star}$ & \textbf{0.166 (0.055)} \\
    FID $\downarrow$   & 375.91 & 100.14 & 127.37 & \textbf{62.28} & 368.58 & 85.79 & 137.83 & \textbf{72.96} \\
    \bottomrule
    \end{tabular}%
    }%
  \label{tab:TABLE_RECON}%
\end{table}%


%% file: TABLE/TABLE_DT_SYSTOLE.tex
\begin{table}[htbp]
  \centering
  \caption{
    Differences of diffusion tensor parameter global mean values between the reference and reconstruction results (undersampled \textit{k}-space zero-filled images ZF included), on systole testing sets Test-S and Test-MI-S.
    Mean absolute error are applied for fractional anisotropy (FA), mean diffusivity (MD),
    and mean absolute angular error are applied for helix angle gradient (HA Slope) and second eigenvector (E2A).
    The results are quoted as `median [interquartile range]'. 
    $^{\star}$ indicates the specific error distribution is significantly different from the \textbf{best results} distribution by Mann-Whitney Test ($p < 0.05$). 
    Data point with \colorbox{bk}{green background} indicates that the specific distribution of corresponding diffusion tensor parameter global mean values is NOT significantly different from the reference distribution by Mann-Whitney Test ($p > 0.05$).
    Units: FA unitless; 
    MD $10^{-3} \cdot \text{mm}^{2}\cdot \text{sec}^{-1}$; 
    HA Slope $\text{degrees} \cdot \text{mm}^{-1}$ 
    and E2A degrees.
  }
    \resizebox{\textwidth}{!}{%
    \begin{tabular}{c|cccc|cccc}
    \toprule
    \textbf{DT Para.} & \multicolumn{4}{c|}{\textbf{Test-S}} & \multicolumn{4}{c}{\textbf{Test-MI-S}} \\
    \midrule
    \textbf{AF $\times 2$} & ZF    & DAGAN & D5C5  & SwinMR & ZF    & DAGAN & D5C5  & SwinMR \\
    \midrule
    FA    & 0.057 [0.021] $^{\star}$ & \cellcolor{bk}0.006 [0.008] & \cellcolor{bk}\textbf{0.005 [0.008]} & \cellcolor{bk}0.008 [0.007] & 0.075 [0.016] $^{\star}$ & \cellcolor{bk}\textbf{0.011 [0.014]} & \cellcolor{bk}0.012 [0.009] & \cellcolor{bk}0.012 [0.007] \\
    MD    & \cellcolor{bk}0.011 [0.012] $^{\star}$ & \cellcolor{bk}0.008 [0.007] $^{\star}$ & \cellcolor{bk}\textbf{0.003 [0.006]} & \cellcolor{bk}\textbf{0.003 [0.009]} & \cellcolor{bk}0.027 [0.014] $^{\star}$ & \cellcolor{bk}0.005 [0.008] & \cellcolor{bk}\textbf{0.002 [0.006]} & \cellcolor{bk}0.004 [0.004] \\
    HA Slope & 1.264 [0.849] $^{\star}$ & \cellcolor{bk}0.244 [0.279] $^{\star}$ & \cellcolor{bk}0.227 [0.228] & \cellcolor{bk}\textbf{0.165 [0.244]} & 1.560 [0.630] $^{\star}$ & \cellcolor{bk}0.220 [0.503] & \cellcolor{bk}0.208 [0.166] & \cellcolor{bk}\textbf{0.164 [0.161]} \\
    E2A & \cellcolor{bk}2.658 [4.213] $^{\star}$ & \cellcolor{bk}0.850 [1.248] $^{\star}$ & \cellcolor{bk}\textbf{0.570 [0.858]} & \cellcolor{bk}0.608 [1.197] & \cellcolor{bk}3.093 [4.065] $^{\star}$ & \cellcolor{bk}0.904 [1.284] & \cellcolor{bk}0.959 [0.993] & \cellcolor{bk}\textbf{0.699 [0.805]} \\
    \midrule
    \textbf{AF $\times 4$} & ZF    & DAGAN & D5C5  & SwinMR & ZF    & DAGAN & D5C5  & SwinMR \\
    \midrule
    FA    & 0.140 [0.043] $^{\star}$ & \cellcolor{bk}\textbf{0.014 [0.020]} & 0.031 [0.028] $^{\star}$ & \cellcolor{bk}0.020 [0.019] & 0.167 [0.043] $^{\star}$ & \cellcolor{bk}\textbf{0.029 [0.022]} & 0.056 [0.020] $^{\star}$ & 0.033 [0.029] \\
    MD    & \cellcolor{bk}0.034 [0.028] $^{\star}$ & \cellcolor{bk}0.022 [0.041] $^{\star}$ & \cellcolor{bk}0.013 [0.020] & \cellcolor{bk}\textbf{0.011 [0.014]} & 0.076 [0.041] $^{\star}$ & \cellcolor{bk}0.026 [0.028] $^{\star}$ & \cellcolor{bk}\textbf{0.009 [0.017]} & \cellcolor{bk}0.013 [0.019] $^{\star}$ \\
    HA Slope & 2.608 [1.080] $^{\star}$ & \cellcolor{bk}0.492 [0.569] & 0.747 [0.728] $^{\star}$ & \cellcolor{bk}\textbf{0.274 [0.433]} & 2.952 [1.084] $^{\star}$ & \cellcolor{bk}0.575 [0.546] & 0.905 [0.820] $^{\star}$ & \cellcolor{bk}\textbf{0.229 [0.368]} \\
    E2A & 10.365 [9.306] $^{\star}$ & \cellcolor{bk}1.977 [2.747] & \cellcolor{bk}\textbf{1.661 [2.874]} & \cellcolor{bk}1.680 [1.557] & 7.372 [6.349] $^{\star}$ & \cellcolor{bk}2.393 [2.551] & \cellcolor{bk}3.444 [3.890] $^{\star}$ & \cellcolor{bk}\textbf{2.372 [1.878]} \\
    \midrule
    \textbf{AF $\times 8$} & ZF    & DAGAN & D5C5  & SwinMR & ZF    & DAGAN & D5C5  & SwinMR \\
    \midrule
    FA    & 0.210 [0.049] $^{\star}$ & \cellcolor{bk}\textbf{0.034 [0.036]} & 0.048 [0.041] $^{\star}$ & 0.045 [0.050] $^{\star}$ & 0.240 [0.037] $^{\star}$ & \textbf{0.051 [0.038]} & 0.122 [0.043] $^{\star}$ & 0.072 [0.053] \\
    MD    & \cellcolor{bk}0.049 [0.045] $^{\star}$ & \cellcolor{bk}0.052 [0.054] $^{\star}$ & \cellcolor{bk}0.039 [0.051] $^{\star}$ & \cellcolor{bk}\textbf{0.031 [0.028]} & 0.139 [0.069] $^{\star}$ & 0.066 [0.071] $^{\star}$ & \cellcolor{bk}\textbf{0.028 [0.035]} & \cellcolor{bk}0.039 [0.057] \\
    HA Slope & 3.839 [1.937] $^{\star}$ & 2.073 [1.495] $^{\star}$ & 2.537 [2.229] $^{\star}$ & \textbf{1.273 [1.611]} & 4.131 [1.124] $^{\star}$ & 2.358 [1.666] & 2.767 [1.496] & \textbf{2.327 [1.973]} \\
    E2A & 14.346 [9.715] $^{\star}$ & 6.359 [8.273] $^{\star}$ & \cellcolor{bk}4.847 [7.203] & \cellcolor{bk}\textbf{2.942 [5.299]} & 11.849 [8.061] $^{\star}$ & 4.916 [9.433] & \cellcolor{bk}5.000 [6.944] & \cellcolor{bk}\textbf{4.334 [7.459]} \\
    \bottomrule
    \end{tabular}%
    }%
  \label{tab:TABLE_DT_SYSTOLE}%
\end{table}%


%% file: TABLE/TABEL_DATASET_DETAIL.tex
\begin{table}[htbp]
  \centering
  \caption{The detailed information of the dataset.
  AMYLOID: amyloidosis;
  DCM: dilated cardiomyopathy;
  rDCM: in-recovery DCM;
  HCM: hypertrophic cardiomyopathy
  HCM G+P-: HCM genotype-positive–phenotype-negative
  MI: acute myocardial infarction.
  }
    \setlength{\tabcolsep}{3mm}{
    \begin{tabular}{cccccccccc}
    \toprule
    \multirow{2}[4]{*}{Disease Type} & \multirow{2}[4]{*}{Cardiac Phase} & \multirow{2}[4]{*}{Total} & \multicolumn{6}{c}{Train \& Validation}       & \multirow{2}[4]{*}{Test} \\
\cmidrule{4-9}          &       &       & All Folds & Fold-1 & Fold-2 & Fold-3 & Fold-4 & Fold-5 &  \\
    \midrule
    Health & Diastole & 112   & 90    & 18    & 18    & 18    & 18    & 18    & 22 \\
    Health & Systole & 129   & 103   & 21    & 21    & 21    & 20    & 20    & 26 \\
    \midrule
    AMYLOID & Diastole & 14    & 11    & 3     & 2     & 2     & 2     & 2     & 3 \\
    AMYLOID & Systole & 17    & 14    & 3     & 3     & 3     & 3     & 2     & 3 \\
    \midrule
    DCM   & Diastole & 23    & 18    & 4     & 4     & 4     & 3     & 3     & 5 \\
    DCM   & Systole & 24    & 19    & 4     & 4     & 4     & 4     & 3     & 5 \\
    \midrule
    HCM (G+P-) & Diastole & 24    & 19    & 4     & 4     & 4     & 4     & 3     & 5 \\
    HCM (G+P-) & Systole & 24    & 19    & 4     & 4     & 4     & 4     & 3     & 5 \\
    \midrule
    HCM   & Diastole & 23    & 18    & 4     & 4     & 4     & 3     & 3     & 5 \\
    HCM   & Systole & 16    & 13    & 3     & 3     & 3     & 2     & 2     & 3 \\
    \midrule
    rDCM  & Diastole & 16    & 13    & 3     & 3     & 3     & 2     & 2     & 3 \\
    rDCM  & Systole & 19    & 15    & 3     & 3     & 3     & 3     & 3     & 4 \\
    \midrule
    MI    & Diastole & 20    & 0     & 0     & 0     & 0     & 0     & 0     & 20 \\
    MI    & Systole & 20    & 0     & 0     & 0     & 0     & 0     & 0     & 20 \\
    \bottomrule
    \end{tabular}%
    }%
  \label{tab:TABEL_DATASET_DETAIL}%
\end{table}%

%% file: TABLE/TABLE_DT_DIASTOLE.tex
\begin{table}[htbp]
  \centering
  \caption{
    Differences of diffusion tensor parameter global mean values between the reference and reconstruction results (undersampled \textit{k}-space zero-filled images ZF included), on diastole testing sets Test-D and Test-MI-D.
    Mean absolute error are applied for fractional anisotropy (FA), mean diffusivity (MD),
    and mean absolute angular error are applied for helix angle gradient (HA Slope) and second eigenvector (E2A).
    The results are quoted as `median [interquartile range]'. 
    $^{\star}$ indicates the specific error distribution is significantly different from the \textbf{best results} distribution by Mann-Whitney Test ($p < 0.05$). 
    Data point with \colorbox{bk}{green background} indicates that the specific distribution of corresponding diffusion tensor parameter global mean values is NOT significantly different from the reference distribution by Mann-Whitney Test ($p > 0.05$).
    Units: FA unitless; 
    MD $10^{-3} \cdot \text{mm}^{2}\cdot \text{sec}^{-1}$; 
    HA Slope $\text{degrees} \cdot \text{mm}^{-1}$ 
    and E2A degrees.
  }
    \resizebox{\textwidth}{!}{%
    \begin{tabular}{c|cccc|cccc}
    \toprule
    \textbf{DT Para.} & \multicolumn{4}{c|}{\textbf{Test-D}} & \multicolumn{4}{c}{\textbf{Test-MI-D}} \\
    \midrule
    \textbf{AF $\times 2$} & ZF    & DAGAN & D5C5  & SwinMR & ZF    & DAGAN & D5C5  & SwinMR \\
    \midrule
    FA    & 0.044 [0.017] $^{\star}$ & \cellcolor{bk}0.007 [0.011] & \cellcolor{bk}0.005 [0.006] & \cellcolor{bk}\textbf{0.004 [0.004]} & 0.058 [0.019] $^{\star}$ & \cellcolor{bk}\textbf{0.007 [0.011]} & \cellcolor{bk}0.010 [0.009] & \cellcolor{bk}0.008 [0.007] \\
    MD    & \cellcolor{bk}0.018 [0.023] $^{\star}$ & \cellcolor{bk}0.013 [0.013] & \cellcolor{bk}\textbf{0.007 [0.010]} & \cellcolor{bk}0.009 [0.009] & \cellcolor{bk}0.042 [0.017] $^{\star}$ & \cellcolor{bk}0.007 [0.012] & \cellcolor{bk}0.007 [0.006] & \cellcolor{bk}\textbf{0.004 [0.006]} \\
    HA Slope & 2.188 [1.105] $^{\star}$ & \cellcolor{bk}0.370 [0.522] $^{\star}$ & \cellcolor{bk}0.238 [0.295] & \cellcolor{bk}\textbf{0.218 [0.377]} & 1.961 [1.065] $^{\star}$ & \cellcolor{bk}0.315 [0.290] & \cellcolor{bk}0.173 [0.256] & \cellcolor{bk}\textbf{0.167 [0.274]} \\
    E2A & 3.124 [3.340] $^{\star}$ & \cellcolor{bk}0.887 [1.259] $^{\star}$ & \cellcolor{bk}0.524 [0.796] & \cellcolor{bk}\textbf{0.497 [0.661]} & \cellcolor{bk}3.609 [3.316] $^{\star}$ & \cellcolor{bk}0.693 [1.246] & \cellcolor{bk}\textbf{0.608 [0.781]} & \cellcolor{bk}0.895 [0.821] \\
    \midrule
    \textbf{AF $\times 4$} & ZF    & DAGAN & D5C5  & SwinMR & ZF    & DAGAN & D5C5  & SwinMR \\
    \midrule
    FA    & 0.120 [0.041] $^{\star}$ & \cellcolor{bk}0.013 [0.017] & \cellcolor{bk}0.013 [0.022] & \cellcolor{bk}\textbf{0.009 [0.012]} & 0.138 [0.024] $^{\star}$ & \cellcolor{bk}0.011 [0.009] & \cellcolor{bk}0.016 [0.018] & \cellcolor{bk}\textbf{0.007 [0.014]} \\
    MD    & 0.075 [0.057] $^{\star}$ & \cellcolor{bk}0.029 [0.030] $^{\star}$ & \cellcolor{bk}\textbf{0.014 [0.022]} & \cellcolor{bk}0.015 [0.014] & 0.130 [0.032] $^{\star}$ & 0.036 [0.031] $^{\star}$ & \cellcolor{bk}\textbf{0.014 [0.018]} & \cellcolor{bk}0.015 [0.017] \\
    HA Slope & 2.783 [1.035] $^{\star}$ & \cellcolor{bk}0.641 [0.977] $^{\star}$ & \cellcolor{bk}0.592 [0.896] $^{\star}$ & \cellcolor{bk}\textbf{0.392 [0.643]} & 3.076 [0.699] $^{\star}$ & \cellcolor{bk}0.508 [0.810] & \cellcolor{bk}0.682 [0.294] $^{\star}$ & \cellcolor{bk}\textbf{0.215 [0.483]} \\
    E2A & 5.093 [3.850] $^{\star}$ & \cellcolor{bk}1.464 [1.922] & \cellcolor{bk}1.422 [1.881] & \cellcolor{bk}\textbf{1.036 [1.200]} & 5.006 [5.062] $^{\star}$ & \cellcolor{bk}1.407 [1.566] & \cellcolor{bk}1.296 [1.195] & \cellcolor{bk}\textbf{1.061 [1.690]} \\
    \midrule
    \textbf{AF $\times 8$} & ZF    & DAGAN & D5C5  & SwinMR & ZF    & DAGAN & D5C5  & SwinMR \\
    \midrule
    FA    & 0.233 [0.080] $^{\star}$ & \cellcolor{bk}0.035 [0.044] & \cellcolor{bk}\textbf{0.030 [0.034]} & \cellcolor{bk}0.033 [0.042] & 0.248 [0.028] $^{\star}$ & \cellcolor{bk}\textbf{0.026 [0.043]} & 0.056 [0.061] & \cellcolor{bk}0.034 [0.038] \\
    MD    & 0.103 [0.094] $^{\star}$ & 0.085 [0.099] $^{\star}$ & 0.068 [0.068] $^{\star}$ & \cellcolor{bk}\textbf{0.041 [0.066]} & 0.232 [0.065] $^{\star}$ & 0.145 [0.067] $^{\star}$ & 0.086 [0.067] $^{\star}$ & \textbf{0.043 [0.069]} \\
    HA Slope & 3.282 [1.301] $^{\star}$ & 3.014 [2.634] $^{\star}$ & 2.145 [1.883] $^{\star}$ & \textbf{1.116 [1.408]} & 3.802 [1.226] $^{\star}$ & 3.101 [1.453] $^{\star}$ & 2.708 [1.591] $^{\star}$ & \textbf{1.599 [1.909]} \\
    E2A & 3.538 [4.984] $^{\star}$ & 4.100 [4.677] & 8.363 [10.243] $^{\star}$ & \cellcolor{bk}\textbf{3.118 [6.069]} & 4.025 [3.634] $^{\star}$ & \cellcolor{bk}\textbf{2.476 [4.193]} & \cellcolor{bk}5.369 [6.679] $^{\star}$ & \cellcolor{bk}3.854 [5.843] \\
    \bottomrule
    \end{tabular}%
    }%
  \label{tab:TABLE_DT_DIASTOLE}%
\end{table}%
